\documentclass[preprint,aps,tightenlines,showpacs]{revtex4}
\usepackage{epsfig}
\usepackage{axodraw}

\newcommand{\beq}{\begin{eqnarray}}
\newcommand{\eeq}{\end{eqnarray}}
\newcommand{\nnb}{\nonumber}

\begin{document}
\preprint{\vbox{\baselineskip16pt
\hbox{AS-ITP-2001-019} 
\hbox{MZ-TH-02-08} 
\hbox{hep-ph/0207179} 
}}
\vspace*{5mm}

\title{Light-Cone HQET Sum Rules for the $B\to\pi$ Transition with 
$1/m_Q$ Corrections}
\author{J. G. K\"orner$^a$, Chun Liu$^{a,\, b}$ and Chi-Tau Yan$^b$}
\affiliation{
$^a$Institut f\"ur Physik, Johannes Gutenberg-Universit\"at, D-55099 Mainz, 
Germany
\\
$^b$Institute of Theoretical Physics, Chinese Academy of Sciences,\\
P.O. Box 2735, Beijing 100080, China}
\date{\today}

\begin{abstract}

The $B\to\pi\ell\nu$ weak decay form factors are calculated via light-cone
sum rules within the framework of the heavy quark effective theory.  We 
calculate the leading and the relevant sub-leading universal form factors. 
Our results are matched to the known soft pion limit. We also address the 
large pion energy limit of our sum rule results.  Our results are compared 
with that of other approaches.  
\end{abstract}


\maketitle

\section{Introduction}

It is well-known that the weak $B\to\pi$ transition is important for the 
extraction of the Cabbibo-Kobayashi-Maskawa matrix element $V_{ub}$ from
semileptonic decays, and for the measurement of CP violation from non-leptonic 
decays.  In order to achieve these goals advances have to be made both on the
experimental and theoretical sides. On the theoretical side an important
task is to reduce the uncertainties in the calculations of the relevant 
hadronic matrix element, represented by
\beq \label{FF}
\langle\pi(p)|\,\bar u\,\gamma^\mu\,b\,|B(P)\rangle &=& f_+(q^2)\,
\bigg[ (P+p)^\mu - {m_B^2-m_\pi^2\over q^2}\,q^\mu \bigg] + f_0(q^2)\,
{m_B^2-m_\pi^2\over q^2}\,q^\mu \,,
\eeq
where $q=P-p$ is the momentum transfer to the leptons.  Up to now, the form 
factors have mostly been calculated via light-cone (LC) sum rules~\cite{lc} in 
full QCD. Instead, here we will work in the context of the heavy quark 
effective theory (HQET)~\cite{hqet}.  The reasons are as follows: 

 (i) Unlike for heavy-to-heavy transitions an application of HQET methods does 
not significantly simplify the analysis for heavy-to-light transitions.  
Nevertheless, the systematic nature of HQET allows one to identify and 
estimate uncertainties in the heavy-to-light transitions more easily.  

 (ii) For an analysis involving the $B$ meson HQET is the correct 
approximation method of QCD.  In certain parts of the phase space of the 
semileptonic decay, the pion is not very energetic so that HQET may still be 
valid.

 (iii) The $B\to\pi\pi$ amplitude has been shown to be factorizable at the 
leading order of the $1/m_b$ expansion \cite{bbns}. Knowledge of the form 
factors based on HQET is needed so as to consistently apply the factorization 
approach.  

 (iv) Furthermore, it is still controversial whether the time-like transition 
$B\to\pi$ at large recoil is governed by perturbative or non-perturbative QCD 
\cite{bbns,kls}.  If non-perturbative effects dominate, then the use of HQET 
for the $B\to\pi$ transitions can be fully justified.

  (v) In addition, the results of HQET can also be applied to $D\to\pi(K)$ 
transitions. In HQET, a model-independent analysis of $B(D)\to\pi$ transitions 
to order $1/m_Q$ has been done in Ref.~\cite{ksm,bpi} and for $B(D)\to\rho$ in 
Ref.~\cite{hly}.  

HQET simplifies the analysis by introducing a set of universal functions.  
However, in order to obtain information on the universal functions themselves, 
some nonperturbative techniques, such as light-cone sum rules or lattice 
simulations, must be used in addition.  Here we adopt the LC sum rule 
method~\cite{lcsr,lcsr1} which is suitable for the calculation of form factors 
when light energetic hadrons are involved.  To our knowledge, the LC-HQET sum 
rules were first applied in Ref.~\cite{dz}.  In Ref.~\cite{ww}, they were used 
at the leading order of HQET for $B\to\pi$ transitions. However, the results 
in this paper differ from those in \cite{ww}.

Note that we distinguish between LC-HQET sum rules and LC-QCD sum rules with 
the $1/m_Q$ expansion~\cite{abs}.  The main reason is that the ways to include 
radiative corrections are different. Another reason is that for $H\to\pi$ 
transitions, there is a subtle difference between the two types of sum rules, 
as will be discussed in the paper.  

In this paper we apply LC-HQET sum rules to the $H\to\pi$ transition to order 
$1/m_Q$. In the next section we give a brief review on the application of HQET 
to the decay $B\to\pi\ell\nu$. In section~\ref{LC-HQET-SR}, the leading and 
next-to-leading order universal functions are calculated by using LC-HQET sum 
rules. In Section~\ref{Numerical} we present our numerical results.  
Section~\ref{Summary} contains a summary and discussion, and a comparison with 
other approaches.

\section{The heavy quark expansion}  

In HQET, the velocity of heavy quark $Q$, $v$, is a well defined quantity.  
The heavy quark field can be represented by the velocity-dependent field, 
\begin{equation}\label{EF}
h_v(x) = \exp({i m_Q\,v\!\cdot\! x)\,P_+\,Q(x) \,,}
\end{equation}
where $\displaystyle P_{+}={1+\rlap/v\over 2}$ projects onto the upper 
component of the heavy quark field $Q(x)$.  To the order $1/m_Q$, the 
effective Lagrangian is given by~\cite{MNrep}
\beq  \label{Eff_Lag}
{\cal{L}}_{\rm HQET} &=& \bar h_v\,i v\!\cdot\!D\,h_v
\,+\, {\frac{1}{2 m_Q}}\,\Big[\, O_{\rm kin}
+ \,C_{\rm mag}(\mu)\, O_{\rm mag} \,\Big] \,,
\eeq
where the gauge-covariant derivative $D_\mu =\partial_\mu -ig_s T^a A^a_\mu$
generates the residual momentum $k_\mu$, and
\beq  \label{operKG}
O_{\rm kin}= \bar h_v\,(i D)^2 h_v \,,  \qquad\quad
O_{\rm mag}= {g_s\over 2}\,\bar h_v\,\sigma_{\alpha\beta}
G^{\alpha\beta} h_v \,.
\eeq
$O_{\rm kin}$ describes the kinetic energy of the heavy quark in the hadron, 
and $O_{\rm mag}$ the heavy quark chromomagnetic energy. The equation of 
motion, $i\, v\cdot D\, h_v=0$, is exactly satisfied. The higher dimension 
operators are treated as perturbative power corrections. In the leading 
logarithmic approximation, the renormalization factor $C_{\rm mag}(\mu)$ is 
given by
\label{running}
\beq
C_{\rm mag}(\mu)={r}(\mu)^{-3}\,,\qquad {r}(\mu)=
\bigg[\, \frac{\alpha(\mu)}{\alpha(m_Q)} \,\bigg]^{\frac{3}{33-2n_f}},
\eeq
where $n_f$ is the number of quarks lighter than the heavy quark $Q$.

In the framework of HQET, it is convenient to work in the matrix 
representation for the description of the hadrons, in which wave functions of 
heavy hadron are only dependent on the heavy quark symmetry  and their Lorentz 
transformation properties. The ground-state pseudo-scalar and vector heavy 
mesons are described by the so-called spin wave function
\beq
{\cal{M}}(v) = P_+\cases{ -\gamma_5\,, & for $J^P=0^-;$ \cr
    \rlap/\epsilon\,, & for $J^P=1^-\,$  \cr}
\eeq
with $\epsilon^\mu$ being the polarization vector.  The form factors are  
considered as functions of the kinematic variable 
\begin{equation}
   v\cdot p = {m_H^2 + m_\pi^2 - q^2\over 2 m_H} \,.
\end{equation} 
Using the mass-independent normalization of the heavy meson state 
$\displaystyle |H(v)\rangle = {m_H^{-1/2}}|H(P)\rangle$, the form factors can 
be re-defined as 
\begin{equation} \label{vpFF}
\langle \pi(p)|\,\bar u\,\gamma_\mu\, Q\,| H(v) \rangle =
2\,[f_1(v\cdot p) \, v_\mu + f_2(v\cdot p) \,\widehat{p}_\mu ]\,,
\end{equation}
where the dimensionless variable is 
$\displaystyle \widehat p^\mu = {p^\mu\over v\cdot p}\,$.  The relation 
between the form factors in Eqs.~(\ref{FF}) and (\ref{vpFF}) is given 
by~\cite{bpi}
\beq \label{FFR}
f_+(q^2) &=& \sqrt{m_H}\,\bigg\{ {f_2(v\cdot p)\over v\cdot p}
+ {f_1(v\cdot p)\over m_H} \bigg\} \,, \\  \label{Scalar}
f_0(q^2) &=& {2\over\sqrt{m_H}}\,{m_H^2\over m_H^2 - m_\pi^2}\,
\bigg\{ \Big[ f_1(v\cdot p) + f_2(v\cdot p) \Big]
- {v\cdot p\over m_H}\,\Big[ f_1(v\cdot p) + \widehat p^2\,
f_2(v\cdot p) \Big] \bigg\} \,.
\eeq

The form factors can be systematically analyzed order by order in terms of 
powers of $1/m_Q$.  Let us begin with the expansion of the heavy-light vector 
current,
\beq
\langle\pi(p)|\bar q\,\gamma^\mu\,Q| H(v)\rangle
&=& \sum_i C_i(\mu)\,\langle\pi(p)|J_i|H(v)\rangle  \nnb \\[-2mm]
&+& \sum_j {1\over 2 m_Q} B_j(\mu) \langle\pi(p)|O_j|H(v)\rangle\, +
{\cal{O}}(\frac{1}{m_Q^2}) \,.
\eeq 
In the limit of massless light quarks, a convenient basis of the above 
operators is~\cite{fnl}
\beq\label{Current}
\begin{array}{ll}
   \,J_1 = \bar q\,\gamma^\mu h_v \,, &\qquad
   \,J_2 = \bar q\,v^\mu h_v \,,
   \\
   O_1 = \bar q\,\gamma^\mu\,i\,\rlap/\!D\,h_v \,, &\qquad
   O_4 = \bar q\,(-i v\!\cdot\!\overleftarrow{D})\,\gamma^\mu h_v \,,
   \\
   O_2 = \bar q\,v^\mu\,i\,\rlap/\!D\,h_v \,, &\qquad
   O_5 = \bar q\,(-i v\!\cdot\!\overleftarrow{D})\,v^\mu h_v \,,
   \\
   O_3 = \bar q\,i D^\mu h_v \,, &\qquad
   O_6 = \bar q\,(-i\overleftarrow{D^\mu})\,h_v \,.
\end{array}
\eeq
The corresponding Wilson coefficients are given by~\cite{fnl,fg}
\beq \begin{array}{ll}
&\displaystyle B_1(\mu) = C_1(\mu) ={r}^{2}\,, \qquad
  B_2(\mu) = {1\over 2}\,B_3(\mu) = C_2(\mu) =0 \,,\\[2mm]
&\displaystyle B_4(\mu) = {34\over 27}\,{r}^{2} - {4\over 27}\,{r}^{-1}
    - {10\over9} + {16\over3}\,{r}^{2}\,\ln {r} \,,  \\[2mm]
&\displaystyle B_5(\mu) = - {28\over27}\,{r}^{2} + {88\over 27}\,{r}^{-1}
 - {20\over9} \,, \qquad B_6(\mu) = - 2\,{r}^{2} - {4\over3}\,{r}^{-1}
 + {10\over3} \,.
\end{array}
\eeq

At the leading order of the $1/m_Q$ expansion, the matrix element of the 
relevant current $\bar q\, \Gamma\, h_v$ can be written as
\beq  \label{LOE}
\langle \pi(p)|\,\bar q\,\Gamma\,h_v\,|H(v)\rangle = - {\rm Tr}\Big\{\,\gamma_5
\Big[\, L_a(v\cdot p,\,\mu) + \rlap/ \widehat p\,L_b(v\cdot p,\,\mu)\, \Big]
\,\Gamma\,{\cal{M}}(v) \,\Big\} \,,
\eeq
where the universal functions $L_{\alpha}(v\cdot p,\mu)~(\alpha =a,\,b)$
depend on the kinematic variable $v\cdot p$, but not on the heavy quark mass
$m_Q$.

At the next-to-leading order in the heavy-quark expansion, the $1/m_Q$
corrections coming from both the effective current and the effective Lagrangian
of HQET will appear as follows. Matrix elements of the operators $O_1$, $O_2$,
and $O_3$ in the effective current can be expressed by the generic
structure 
\beq \label{ECC}
&&\langle \pi(p)|\,\bar q\,(\Gamma\,i D)_\mu h_v\,|H(v)\rangle \nnb \\
&&= - {\rm Tr}\Big\{\,
\Big[\,( F_1\,v_\mu + F_2\,\widehat p_\mu + F_3\,\gamma_\mu)\,\gamma_5\,
+\,( F_4\,v_\mu + F_5\,\widehat p_\mu + F_6\,\gamma_\mu)\,\gamma_5\,
\,\rlap/\widehat p\,\Big]\,\Gamma\,{\cal{M}}(v) \,\Big\} \,,
\eeq
where the universal functions $F_i(v\!\cdot\! p,\mu)~(i=1,\cdots,6)$ are also
$m_Q$-independent.
Matrix elements of the operators $O_4$, $O_5$, and $O_6$ are not independent, 
and can be obtained from the above structures Eqs.(\ref{LOE}) and (\ref{ECC}).
Additionally, corrections coming from insertions of the operators 
$O_{\rm kin}$ and $O_{\rm mag}$ into matrix elements of the leading-order 
currents can be described by six additional universal functions
$K_{\alpha}(v\cdot p,\mu)~(\alpha =a,\,b)$ and 
$S_i(v\cdot p,\mu)~(i=1,\cdots, 4)$, which are defined by matrix elements of 
the time-ordered products
\beq
&&\langle \pi(p)|\,i\!\int\! {\rm d}^4y\,T\Big\{ \bar q\,\Gamma\,h_v(0),
O_{\rm kin}(y) \Big\}\,|H(v)\rangle = - {\rm Tr}\Big\{\, \gamma_5\,
(\,K_a + \,\rlap/\widehat p\,K_b\,)\,\Gamma\,{\cal{M}}(v) \,\Big\} \,,
\label{Ki}
\\[3mm]
\label{Si}
&&\langle\pi(p)|\,i\!\int\! {\rm d}^4y\,T\Big\{ \bar q\,\Gamma\,h_v(0),
O_{\rm mag}(y) \Big\}\,|H(v)\rangle \nonumber\\
&&= - {\rm Tr}\bigg\{\, \Big[\, (i S_1\,\widehat p_\alpha
    \gamma_\beta + S_2\,\sigma_{\alpha\beta})\,\gamma_5
    + (i S_3\,\widehat p_\alpha\gamma_\beta
    + S_4\,\sigma_{\alpha\beta})\,\gamma_5\,
    \,\rlap/\widehat p\,\Big]\,\Gamma\,P_+\,
    \sigma^{\alpha\beta}\,{\cal{M}}(v) \,\bigg\}\,.
\eeq

The form factors in Eq.~(\ref{vpFF}) can then be expanded as follows after 
doing the appropriate traces~\cite{bpi},
\beq \label{FormFactors}
f_1 &=& C_1\,L_a + {1\over 2 m_Q}\,\bigg\{\,C_1\,\Big[\, {\cal F}_a^1 + ( K_a 
    +C_{\rm mag}\,{\cal S}_a) \,\Big] - B_4\,(\bar\Lambda -v\cdot p)\,L_a
    \,-\, B_5\,(\bar\Lambda -v\cdot p) \nnb \\[-2mm]
    & &\qquad\qquad\qquad\qquad \times \,(L_a+L_b)
    \,+\, B_6\Big[\,{\cal F}_a^3 - \bar\Lambda(L_a+L_b)\,\Big] \;\bigg\}
    \,+\, {\cal{O}}(\frac{1}{m_Q^2})\,,   \\[0mm]
f_2 &=& C_1\,L_b + {1\over 2 m_Q}\,\bigg\{\,C_1\,\Big[\, {\cal F}_b^1 + ( K_b 
    +C_{\rm mag}\,{\cal S}_b ) \,\Big] - B_4\,(\bar\Lambda -v\cdot p)\,L_b  
    \nnb \\[-2mm]
    & &\qquad\qquad\qquad\qquad\qquad\qquad
    \,+\, B_6\,\Big[\,{\cal F}_b^3 + v\cdot p \,(L_a+L_b) \,\Big]\;\bigg\}
    \,+\, {\cal{O}}(\frac{1}{m_Q^2})\,,
\eeq
where ${\cal F}_{\alpha}^i$ ($i=1,\,3$) and ${\cal S}_{\alpha}$
are defined by
\beq \label{FSdef}
\begin{array}{ll}\label{CC}
{\cal F}_a^1= F_1+2F_3-\widehat p^2 F_5, &\quad
{\cal F}_a^3= F_1-F_3+F_4 \,,
\\
{\cal F}_b^1= F_2+F_4+2F_5-4F_6,  &\quad
{\cal F}_b^3= F_2+F_5-F_6 \,,
\\[0.5mm]
{\cal S}_a= - 2 S_1 + 6 S_2 + 2\widehat p^2\,S_3\,, &\quad
{\cal S}_b= 2 S_1 - 2 S_3 + 6 S_4\,.
\end{array}
\eeq
It should be mentioned that consequences of the heavy quark symmetry and
the equations of motion for the heavy and light quark fields
make only two ($F_5$ and $F_6$) of the six form factors $F_i$'s 
independent~\cite{bpi}. This means that one can
re-express ${\cal F}_{\alpha}^i$ by $L_a,\,L_b$ and $F_6$ as follows
\beq \label{consEM}
\begin{array}{ll}
{\cal F}_a^1 = - (\bar\Lambda-2v\cdot p)L_a + v\cdot p\,\widehat p^2L_b
                + 4F_6\,, & \qquad
{\cal F}_a^3 =   v\cdot p\,L_a + \bar\Lambda L_b  + 2F_6\,, \\
{\cal F}_b^1 = - v\cdot p\,L_a - \bar\Lambda L_b  - 4F_6\,, & \qquad
{\cal F}_b^3 = - v\cdot p\,L_a - \bar\Lambda L_b  - 2F_6\,.
\end{array}
\eeq
Note that between the two independent universal functions from the effective 
current corrections only $F_6(v\!\cdot\! p,\mu)$ is relevant to the $H\to\pi$ 
matrix element in Eq.~(\ref{FF})
\beq \label{Fsix}
F_6&=& - \frac{1}{2}\,\Big(\,{\cal F}_b^1 + {\cal F}_a^3 \,\Big)\,.
\eeq

\section{The light-cone HQET sum rules} \label{LC-HQET-SR}

Our aim is to calculate the independent form factors given in the last section
by LC-HQET sum rules. In subsection A, the leading universal functions
$L_\alpha$ ($\alpha =a,\,b$) are calculated. The relevant sub-leading universal
functions, $F_6$ and $K_\alpha + C_{\rm mag}{\cal S}_\alpha$, will be 
calculated in subsections B and C, respectively.

\subsection{Leading order}

Let us begin with the following 2-point vacuum-pion correlation function as
depicted in Fig.~\ref{FigLO}:
\vspace*{2mm}
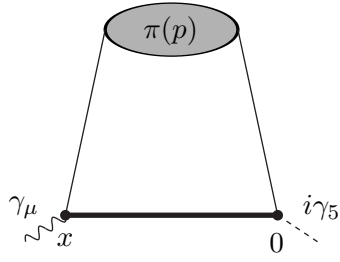
\begin{figure}
\begin{center}
\begin{picture}(140,100)(-5,0)
\GOval(70,85)(10,25)(0){0.7}
\Text(70,85)[]{$\pi (p)$}
\SetWidth{2.0}
\Line(30,15)(110,15)
\SetWidth{0.5}
\Line(30,15)(45,85)
\Line(110,15)(95,85)
\Photon(30,15)(15,5){2}{3}
\DashLine(110,15)(125,5){2}
\Vertex(30,15){2}
\Vertex(110,15){2}
\Text(30,8)[t]{$x$}
\Text(110,8)[t]{$0$}
\Text(20,18)[r]{$\gamma_\mu$}
\Text(120,18)[l]{$i\gamma_5$}
\end{picture}
\end{center}
\caption{\it The diagrammatic representation of the 2-point vacuum-pion 
correlation function Eq.~(\ref{CFLO}), where the heavy solid line represents 
the heavy quark.
\label{FigLO}}
\end{figure}
\beq \label{CFLO}
F_\mu (\lambda,\omega)&=&i\!\int\! {\rm d}^4 x \,{\rm e}^{i k\cdot x}
\,\langle \pi(p)|\,T\Big\{\bar u(x)\gamma_\mu h_v(x), \,\bar h_v(0)
i \gamma_5 d(0) \Big\} \,| 0 \rangle \   \nnb \\
& = &F_a^{\rm}(\lambda, \omega)\,v_\mu\,+
    \,F_b^{\rm}(\lambda, \omega)\,\widehat p_\mu\,,
\eeq
where $\lambda=2v\cdot p$ and $\omega=2v\cdot k$. On the one hand, by 
inserting a complete set of intermediate states with the same quantum numbers 
as the $H$ meson between the current in the above vacuum-pion correlation
function, we obtain the hadronic representation
\beq
F^{\rm Hadr}_{\alpha}(\lambda, \tilde{\omega})&=&
2F\,\frac{L_{\alpha}(\lambda)}{2\bar\Lambda-\tilde{\omega}}\,+\,{resonances}\,,
\label{7}
\eeq
where
$\tilde{\omega} = 2 v\cdot \tilde k$ with $\tilde k$ being $k +p$.
$\bar{\Lambda}$ is the heavy meson mass as defined in the HQET.
{\it Note that it is $\tilde{\omega}$ that is the relevant energy of the
quark system}. The decay constant $F$ is defined as
\beq
\langle H(v) | \bar h_v  \Gamma d| 0\rangle &=& \frac{i}{2}F(\mu){\rm Tr}
\Big[ \Gamma {\cal M}(v) \Big]\,,
\eeq
with $\Gamma$ being an arbitrary Dirac matrix.

On the other hand the correlation function can be calculated by
expanding the $T$-product of the currents near the light-cone at $x^2=0$.
In our calculations, the chiral limit $p^2=m_\pi^2=0$ is taken.
By adopting the fixed-point gauge in which
\beq \label{fix-point}
x_\mu A^\mu(x)=0\,,
\eeq
the following complete heavy quark propagator equals the free one,
\beq
\langle 0|T\big\{\, h_v(x)\,\, \bar h_v(0)\,\big\}|0\rangle &=&\int^{\infty}_0
\!{\rm d}\, t\,\delta (x-vt)\frac{1+\rlap/ v}{2}\,.
\label{qp}
\eeq
The matrix elements of the nonlocal vacuum-to-pion operators can be 
parameterized by pion wave functions on the light-cone.  The LC wave functions 
are classified according to their twist and the number of partons they
describe. Here, let ${\cal WF}(n)$ denote the $n$-particle light-cone
wave function.  
One can easily find that at the leading order of HQET only the ${\cal WF}(2)$s 
are relevant. Up to twist four, the matrix elements are parameterized as
follows~\cite{BF, lcsr}:
\beq  \label{2pWF}
\langle\pi(p)|\bar{u}(x)\gamma_\mu\gamma_5d(0)|0\rangle &=&-if_\pi p_\mu
\int_0^1{\rm d}u\,e^{iu\,px}\Big(\varphi_\pi(u)+x^2g_1(u)\Big) \nnb \\[3mm]
& &+f_\pi\Big(x_\mu -\frac{x^2p_\mu}{px}\Big)\int_0^1 {\rm d}u\,
e^{iu\,px}g_2(u)\,,\nnb \\[3mm]
\langle\pi(p)|\bar{u}(x)i\gamma_5d(0)|0\rangle &=&
f_\pi\mu_\pi\int_0^1{\rm d}u~e^{iu\,px}\varphi_{p}(u)\,,\nnb \\[3mm]
\langle\pi(p)|\bar{u}(x)\sigma_{\mu\nu}\gamma_5d(0)|0\rangle &=&i\,
\frac{f_\pi\mu_\pi}{6}\,\Big(p_\mu x_\nu -p_\nu x_\mu \Big)
\int_0^1{\rm d}u~e^{iu\,px}\varphi_{\sigma}(u)\,,
\eeq
with $\mu_\pi=m_\pi^2/(m_u + m_d)$.  Here, $\varphi_\pi$ is the leading
twist-2 wave function, $g_1$  and $g_2$ are twist-4 wave functions and
$\varphi_p, \varphi_\sigma$  twist-3 wave functions.  In the fixed-point gauge,
the operator $\Pi_G=P\exp\Big\{ig_s\int_0^1d\alpha~x_\mu A^\mu
(\alpha x)\Big\}$ is unity. Considering the identity
\begin{equation} \label{Id1}
\gamma_\mu\gamma_\nu=-i\sigma_{\mu\nu} + g_{\mu\nu}\,,
\end{equation}
one can obtain $F_{\alpha}^{\scriptscriptstyle\rm HQET}(\lambda, \tilde\omega)$
in terms of the light-cone wave functions:
\beq
F_{a}^{\scriptscriptstyle \rm LO}(\lambda, \tilde\omega) &=&
\frac{f_\pi}{2}\! \int^1_0\! {\rm d}u \int^{\infty}_{0}\! i\,{\rm d}t\,
{\rm e}^{i\frac{\tilde\omega}{2}t}\,
\Big[ \,\mu_{\pi}\varphi_p(u) - \Big(\frac{\mu_{\pi}}{6}\varphi_\sigma(u)-
\frac{2}{\lambda}g_2(u) \Big)\frac{\rm d\,}{{\rm d}u}\, \Big]
{\rm e}^{-i\bar u\frac{\lambda}{2}t}\,, \\[3mm]
F_{b}^{\scriptscriptstyle \rm LO}(\lambda, \tilde\omega) &=&
\frac{f_\pi}{2}\! \int^1_0\! {\rm d}u \int^{\infty}_{0}\! i\,{\rm d}t\,
{\rm e}^{i\frac{\tilde\omega}{2}t}\,
\Big[ \,\frac{\lambda}{2}\varphi_{\pi}(u) + \Big( \frac{\mu_{\pi}}{6}
\varphi_{\sigma}(u)
-\frac{2}{\lambda}g_2(u) \Big) \frac{\rm d\,}{{\rm d}u}\, \nnb \\
&&\qquad\qquad\quad\qquad\qquad \quad
- \frac{2}{\lambda}g_1(u) \frac{{\rm d}^2}{{\rm d}u^2}\,\Big]
{\rm e}^{-i\bar u\frac{\lambda}{2}t}\,,
\eeq
where here and below $\bar u=1-u$, and the superscript $ LO $ denotes the
leading order in the $1/m_Q$ expansion.

Setting $F^{\rm Hadr}_{\alpha}$ equal 
$F^{\scriptscriptstyle\rm HQET}_{\alpha}$ defines the LC-HQET sum rules. The 
quark-hadron duality assumption is used to substitute the unknown $resonances$ 
by the HQET result after a dispersion integration above some given threshold 
$\omega_c$,
\beq
resonances &=& \frac{1}{\pi}\int_{\omega_c}^\infty {\rm d} \nu\frac{{\rm Im}
F_{\alpha}^{\scriptscriptstyle\rm HQET}(\lambda, \nu)}{\nu-\tilde{\omega}}
\,+\,subtraction\,.
\eeq
The HQET spectral density can be obtained by the following double Borel 
transformation,
\beq
\frac{1}{\pi}{\rm Im}F_{\alpha}^{\scriptscriptstyle\rm HQET}(\lambda,\nu)&=&
\widehat{B}^{(-\frac{1}{\tau})}_{\frac{1}{\nu}}\,
\widehat{B}^{(\tilde{\omega})}_{\tau}\,F_{\alpha}^{\scriptscriptstyle\rm HQET}
(\lambda,\tilde{\omega}) 
\eeq 
with the Borel transformation being defined as 
\beq
\widehat{B}^{(X)}_Y&=&\lim_{ {\scriptstyle
\stackrel{X\to\infty,}{n\to\infty,}\,\,} \stackrel{Y=\frac{-X}{n} {\rm fixed}}
{}}Y\frac{(-X)^n}{\Gamma(n)}\frac{{\rm d}^n\,}{{\rm d}X^n}\,
\eeq
whose property,
$\displaystyle B^{(\tilde\omega)}_{\tau}{\rm e}^{\,\rho\,\tilde\omega}=
\delta (\rho-1/{\tau})$, is very useful in the calculation here.
Evidently, before doing Borel transformations, one should first perform
the Wick rotation on $t$.

Finally, by performing a Borel transformation $\widehat B_T^{\,(\tilde{\omega})}$
on both sides of the sum rule, so as to enhance the ground state contribution,
suppress higher twist terms and remove the $subtraction$, one can obtain the 
sum rules,
\beq
L_{\alpha}(\lambda)&=&
\frac{1}{2F}\,{\rm e}^{2\bar\Lambda/T}\int^{\omega_c}_0 {\rm d}\,\nu\,
\frac{1}{\pi}\,{\rm Im}
F_{\alpha}^{\scriptscriptstyle\rm LO}(\lambda, \nu)\,
{\rm e}^{-\nu/T}, 
\label{SR}
\eeq
where $T$ is the Borel parameter, and  
\beq
\frac{1}{\pi}{\rm Im}F_a^{\scriptscriptstyle\rm LO}(\lambda, \nu)
& = & \frac{f_\pi}{2}\,\Theta (u_0)\,\Big[\,\frac{2\mu_\pi}{\lambda}\varphi_p
(u_0)+\frac{\mu_\pi}{3\lambda}\varphi'_\sigma (u_0)
-\frac{4}{\lambda^2}g_2'(u_0)\,\Big]\,,
\\[3mm]
\frac{1}{\pi}{\rm Im}F_b^{\scriptscriptstyle\rm LO}(\lambda, \nu)\,
&=&\frac{f_{\pi}}{2}\,\Theta (u_0)\,\Big[\,\varphi_\pi(u_0)-
\frac{\mu_\pi}{3\lambda}\varphi'_\sigma (u_0) -\frac{4}{\lambda^2}g_1''(u_0)+
\frac{4}{\lambda^2}g_2'(u_0)\,\Big]\,.
\eeq
Here and in the following, $u_0 = 1-\frac{\nu}{\lambda}$, and the prime 
denotes derivatives. After integration by parts, we obtain the final sum rules,
\beq
L_a(v\cdot p)&=&\,\frac{f_{\pi}}{2\,F\,} \,{\rm e}^{2\bar\Lambda/T}
\Bigg\{ \, \int^\theta_0 {\rm d} u\,\Big[\,\mu_\pi\varphi_p(\bar u) -
\frac{\mu_\pi v\cdot p}{3T}\varphi_\sigma(\bar u) +
\frac{2}{T}g_2(\bar u)\,\Big]\, {\rm e}^{-2u\,v\cdot p/T} \nnb \\
&&\quad\quad\quad\quad\quad\,
-\Big[\,\frac{\mu_\pi}{6}\varphi_\sigma(\bar\theta)-\frac{1}{v\!\cdot\! p}
g_2(\bar \theta)\,\Big]\, {\rm e}^{-2\theta\,v\cdot p/T}\; \Bigg\}\, ,
\label{sumruleA}
\eeq
\beq
L_b(v\cdot p)&=&\frac{f_{\pi}}{2\,F\,} \, {\rm e}^{2\bar\Lambda/T}
\,\Bigg\{\,v\!\cdot\! p\,\int^\theta_0 {\rm d} u\,\Big[\,\varphi_\pi(\bar u) +
\frac{\mu_\pi}{3T}\varphi_\sigma(\bar u) - \frac{4}{T^2}g_1(\bar u)-
\frac{2}{v\cdot p\,T} g_2(\bar u) \Big]
 {\rm e}^{-2u\,v\cdot p/T} \nnb \\
& & \quad\quad\quad\quad\quad +
\Big[\,\frac{\mu_\pi}{6}\varphi_\sigma(\bar\theta)-\frac{2}{T}g_1(\bar\theta)+
\frac{1}{v\!\cdot \!p}\Big(\frac{{\rm d}g_1(\bar\theta)}{{\rm d}u}
-g_2(\bar\theta)\Big)\,\Big]\,{\rm e}^{-2\theta\,v\cdot p/T} \; \Bigg\} \,,
\label{sumruleB}
\eeq
where $\theta={\rm Min}(1, \frac{\omega_c}{2v\cdot p})$ and
$\bar\theta = 1- \theta$. We see that the LC sum rules become meaningless for 
processes with a very soft pion which would enhance higher twist contributions.

\subsection{$1/m_Q$ corrections from the effective current}

To obtain the $1/m_Q$ corrections from the effective current, we consider
the 2-point vacuum-pion correlation function containing a
covariant derivative, shown in Fig.~\ref{Fig_EC}:
\vspace*{2mm}
\begin{figure}
\begin{center}
\begin{picture}(320,100)(-5,0)
\Text(70,-3)[t]{(a)}
\GOval(70,85)(10,25)(0){0.7}
\Text(70,85)[]{$\pi (p)$}
\SetWidth{2.0}
\Line(30,15)(110,15)
\SetWidth{0.5}
\Line(30,15)(45,85)
\Line(110,15)(95,85)
\Photon(30,15)(15,5){2}{3}
\DashLine(110,15)(125,5){2}
\GOval(30,15)(2.2,4.4)(0){0.0}
\Vertex(110,15){2}
\Text(30,8)[t]{$x$}
\Text(110,8)[t]{$0$}
\Text(27,22)[r]{${}_{(i \Gamma^m \partial )_\mu}$}
\Text(115,18)[l]{$i\gamma_5$}
\Text(250,-3)[t]{(b)}
\GOval(250,85)(10,25)(0){0.7}
\Text(250,85)[]{$\pi (p)$}
\SetWidth{2.0}
\Line(210,15)(290,15)
\SetWidth{0.5}
\Line(210,15)(225,85)
\Line(290,15)(275,85)
\Photon(210,15)(195,5){2}{3}
\DashLine(290,15)(305,5){2}
\Vertex(210,15){2}
\Vertex(290,15){2}
\Text(210,8)[t]{$x$}
\Text(290,8)[t]{$0$}
\Text(208,23)[r]{${}_{(-i \Gamma^m g_s A)_\mu}$}
\Text(295,18)[l]{$i\gamma_5$}
\Gluon(210,15)(250,75){-3}{7}
\end{picture}
\end{center}
\caption{\it The diagrammatic representation of the 2-point vacuum-pion 
correlation function Eq.~(\ref{CF_EC}), where the heavy solid lines represent 
the heavy quark, and the helical lines a gluon.
\label{Fig_EC}}
\end{figure}
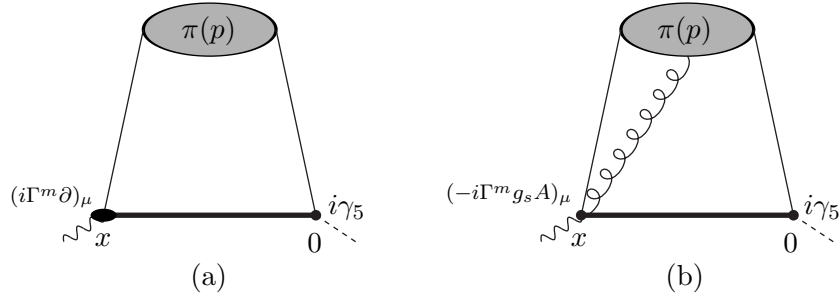
\beq \label{CF_EC}
{E}_{\mu}^m (\lambda,\tilde \omega) &=& i\!\int\! {\rm d}^4 x\,
{\rm e}^{ik\cdot x}\,
\langle \pi(p)|\,T\Big\{ \bar u(x) \,(\Gamma^m i D)_\mu \,h_v(x), \,
\bar h_v(0) i \gamma_5 d(0) \Big\} \,| 0 \rangle  \nnb \\
&=&{E}_a^m (\lambda,\tilde \omega)v_\mu +
{E}_b^m (\lambda,\tilde \omega) \widehat p_\mu \,,
\eeq
where $m=1,\,3$ is an index introduced for convenience, and
\beq
(\Gamma^{1}i D)_{\mu}=\gamma_\mu \gamma_\beta  i D^\beta\,,  \qquad
(\Gamma^{3}i D)_{\mu}=i D_\mu\,.
\eeq
The hadronic representation of this 2-point correlation function
can be expressed as
\beq
E^{\rm m\,Hadr}_{\alpha}(\lambda, \tilde{\omega})&=&
2 F\,\frac{{\cal F}^m_{\alpha}(\lambda)}{2\bar\Lambda-\tilde{\omega}}
\,+\,{resonances}\,,
\eeq
where the ${\cal F}^m_{\alpha}(\lambda)$ have been defined in Eq.(\ref{FSdef}).

The above correlation function can be calculated directly in HQET.
We choose the relevant momentum $\tilde k$ to be parallel to $v$, {\it i.e.},
$\tilde k_\mu =(\tilde\omega/2)\, v_\mu $, and work in the fixed-point gauge
where $A_\mu(x)$ can be expressed as
\beq \label{A_fixed_point}
A_\beta (x)&=&x^\alpha\int^1_0 {\rm d}w\,w\,G_{\alpha\beta}(w x)\,.
\eeq
Eq.(\ref{CF_EC}) then can be given in terms of the above ${\cal WF}(2)$s,
and the following $3$-particle wave functions~\cite{BF,lcsr}:
\beq \label{3pWF}
&&\langle\pi(p)|\bar{u}(x)\sigma_{\mu\nu}\gamma_5g_s G_{\alpha\beta}(wx)d(0)
|0\rangle \nonumber \\
&&\quad =if_{3\pi}\Big[(p_\alpha p_\mu g_{\beta\nu}-p_\beta p_\mu 
g_{\alpha\nu})-(p_\alpha p_\nu g_{\beta\mu}-p_\beta p_\nu g_{\alpha\mu})\Big]
\int{\cal D}\alpha_i\,\varphi_{3\pi}(\alpha_i)e^{i(\alpha_1+w\alpha_3)\,px}\,,
\nnb \\[3mm]
&&\langle\pi(p)|\bar{u}(x)\gamma_\mu\gamma_5 g_sG_{\alpha\beta}(wx)d(0)
|0\rangle \nonumber \\
&&\quad =f_\pi\bigg[ p_\beta\Big( g_{\alpha\mu}-\frac{x_\alpha p_\mu}{px}\Big) 
-p_\alpha\Big(g_{\beta\mu}-\frac{x_\beta p_\mu}{px}\Big)\bigg]
\int{\cal D}\alpha_i\varphi_\perp (\alpha_i)e^{i(\alpha_1+w\alpha_3)\,px}
\nonumber \\
&& \qquad {}+f_\pi\frac{p_\mu}{px}(p_\alpha x_\beta -p_\beta x_\alpha )
\int{\cal D}\alpha_i\,\varphi_\parallel (\alpha_i)e^{i(\alpha_1+w\alpha_3)\,
px}\,, \nnb \\[3mm]
&&\langle\pi(p)|\bar{u}(x)\gamma_\mu g_s\tilde{G}_{\alpha\beta}(wx)d(0)
|0\rangle \nonumber\\
&&\quad =if_\pi\bigg[ p_\beta\Big( g_{\alpha\mu}-\frac{x_\alpha p_\mu}{px}\Big)
-p_\alpha\Big( g_{\beta\mu}-\frac{x_\beta p_\mu}{px}\Big)\bigg]
\int{\cal D}\alpha_i\,\tilde{\varphi}_\perp (\alpha_i)e^{i(\alpha_1+w\alpha_3)
\,px} \nonumber \\
&&\qquad {}+if_\pi\frac{p_\mu}{px}(p_\alpha x_\beta -p_\beta x_\alpha )
\int{\cal D}\alpha_i\,\tilde{\varphi}_\parallel (\alpha_i)
e^{i(\alpha_1+w\alpha_3)\,px}\,,
\eeq
where $\tilde{G}_{\alpha\beta}=\frac{1}{2} \epsilon _{\alpha\beta \sigma\tau}
G^{\sigma\tau}$, and ${\cal D}\alpha_i= d\alpha_1 d\alpha_2 d\alpha_3
\delta(1-\alpha_1-\alpha_2-\alpha_3)$.
The wave function
$\varphi_{3\pi}(\alpha_i)=\varphi_{3\pi}(\alpha_1,\alpha_2,\alpha_3)$
is twist 3, and $\varphi_\perp$, $\varphi_\parallel$,
$\tilde{\varphi}_\perp$ and $\tilde{\varphi}_\parallel$ are twist 4.
Using the identities, Eq.~(\ref{Id1}) and
\beq \label{Id2}
\gamma_\mu \gamma_\alpha \gamma_\beta = \gamma_\mu g_{\alpha\beta}
-\gamma_\alpha g_{\mu\beta} + \gamma_\beta g_{\mu\alpha}
- i \,\varepsilon_{\mu\alpha\beta\delta} \gamma^\delta \gamma_5\,,
\eeq
one obtains
\beq
E_\mu^{1}(\lambda,\tilde\omega)
&=&\frac{f_\pi}{2}\int_0^1 {\rm d} u\int^\infty_0 {\rm d}t \,
{\rm e}^{i\frac{\tilde\omega}{2}t}\,\bigg\{\,
{\frac{\scriptstyle\overleftarrow{\rm d}}{{\rm d} t}}
\bigg(\, \frac{\lambda}{2}\Big[\,\varphi_\pi(u)+t^2g_1(u)\,\Big]\widehat p_\mu
+\mu_\pi\varphi_p(u)v_{\mu} \nnb \\
&& + \,i t\Big[\,\frac{\mu_{\pi}}{12}\lambda\varphi_{\sigma}(u)-g_2(u)\,\Big]
(\widehat p_\mu -v_\mu)\,\bigg) - i\bar u\,\frac{\lambda}{2}\,\widehat p_\mu\,
\bigg(\varphi_\pi(u)+t^2g_1(u)  \nnb \\
&& +\mu_\pi\varphi_p(u)
+ i t \Big[ \,\frac{\mu_{\pi}}{12}\lambda\varphi_{\sigma}(u)-g_2(u) \,\Big]\,
\bigg) + i \lambda \Big[\,\frac{\mu_\pi}{4}\varphi_\sigma(u)
 - i t g_1(u)\,\Big]\widehat p_\mu  \nnb \\
&& - 2ig_2(u) v_\mu \,\bigg\}\:{\rm e}^{-i\bar u\frac{\lambda}{2}t } \,
-\frac{f_\pi}{2}\int\! {\cal D} \alpha_i \!
\int_0^1\! {\rm d}w\,w \int^\infty_0\! {\rm d}t \,
{\rm e}^{i\frac{\tilde\omega}{2}t} \,\frac{\lambda}{2}\,t\,\bigg\{\,
\Big[\,\lambda \frac{f_{3\pi}}{f_\pi}\varphi_{3\pi}(\alpha_i) \nnb \\
&& -\,\varphi_{\parallel}(\alpha_i)\,\Big] \widehat p_\mu
+ 2\varphi_{\perp}(\alpha_i)v_\mu
\,\bigg\}\,{\rm e}^{-i(1-\alpha_1-w\alpha_3)\frac{\lambda}{2}t}\,, \quad
\\[2mm]
{E}_\mu^{3}(\lambda,\tilde \omega)
&=&\frac{f_\pi}{2}\int_0^1 {\rm d} u \int^\infty_0 {\rm d}t \,
{\rm e}^{i\frac{\tilde\omega}{2}t}\,\bigg\{\,
{\frac{\scriptstyle\overleftarrow{\rm d}}{{\rm d} t}}
\bigg(\,\frac{\lambda}{2}\Big[\,\varphi_\pi(u)+t^2g_1(u)\,\Big]
+\mu_\pi\varphi_p(u)\,\bigg) v_{\mu} \nnb \\
&& -\, i\bar u\,\frac{\lambda}{2}\,\widehat p_\mu\,
\bigg(\,\frac{\lambda}{2}\Big[\,\varphi_\pi(u)+t^2g_1(u)\,\Big]
+\mu_\pi\varphi_p(u)\,\bigg) +\lambda t g_1(u)v_\mu \nnb \\
&&+\,ig_2(u)(\widehat p_\mu -v_\mu)\,\bigg\}\:{\rm e}^{-i\bar u
\frac{\lambda}{2}t }\, -\frac{f_\pi}{2}\int\! {\cal D} \alpha_i \!
\int_0^1\! {\rm d} w\,w\int^\infty_0\! {\rm d} t \,
{\rm e}^{i\frac{\tilde\omega}{2}t }\,\frac{\lambda}{2}\,t\, \nnb \\
&& \times\,(v_\mu - \widehat p_\mu)\,\varphi_{\parallel}(\alpha_i)\,
{\rm e}^{-i(1-\alpha_1-w\alpha_3)\frac{\lambda}{2}t }\,,
\eeq
where terms proportional to $\widehat p^2$ are omitted in the chiral limit.

After performing the corresponding Borel transformations, we obtain the sum
rules for the effective current corrections:
\beq
{\cal F}^m_{\alpha}(\lambda)\, = \,\frac{\,{\rm e}^{2\bar\Lambda/T}}{2F}
\int^{\omega_c}_0 {\rm d}\,\nu\,\frac{1}{\pi}\,{\rm Im} {E}^{m}_{\alpha}
(\lambda, \nu)\,{\rm e}^{-\nu/T}\,,
\eeq
where
\beq
\frac{1}{\pi}{\rm Im}{E}_a^1(\lambda,\nu)&=&
\frac{f_\pi}{2}\Theta(u_0)\Big[\,\bar u_0 \Big(\,\mu_\pi\varphi_{p}(u_0)
+\frac{\mu_\pi}{6}\varphi'_{\sigma}(u_0) - \frac{2}{\lambda}g'_2(u_0)\,\Big)
-\frac{4}{\lambda}g_2(u_0) \nnb \\
&& \qquad\qquad -\,\frac{2}{\lambda}
\int^{u_0}_0\! d \alpha_1 \int^{1-\alpha_1}_{u_0-\alpha_1}
d\alpha_3\,\frac{1}{\alpha_3^2}\,2\,\varphi_{\perp}(\alpha_1,1-\alpha_1-
\alpha_3, \alpha_3)\,\Big]\,,
\\[3mm]
\frac{1}{\pi}{\rm Im}{E}_b^1(\lambda,\nu)&=&
\frac{f_\pi}{2}\Theta(u_0)\Big[ -\bar u_0 \Big(\,\frac{\lambda}{2}
\varphi_\pi(u_0)+\mu_\pi\varphi_{p}(u_0) - \frac{2}{\lambda}g''_1(u_0)\,\Big)
+\frac{\mu_\pi}{3}\varphi_\sigma(u_0) \nnb \\
&& +\frac{2}{\lambda}g_2(u_0) -\frac{4}{\lambda}g'_1(u_0)
+\,\frac{2}{\lambda}\int^{u_0}_0\! d \alpha_1 \int^{1-\alpha_1}_{u_0-\alpha_1}
d\alpha_3\,\frac{1}{\alpha_3^2}\,  \nnb \\
&&\times \Big(\,\varphi_{\parallel}(\alpha_1,1-\alpha_1-\alpha_3,\alpha_3)
-\lambda\frac{f_{3\pi}}{f_\pi}\varphi_{3\pi}(\alpha_1,1-\alpha_1-\alpha_3,
\alpha_3)\, \Big)\,\Big]\,,
\\[3mm]
\frac{1}{\pi}{\rm Im}{E}_a^3(\lambda,\nu)&=&
\frac{f_\pi}{2}\Theta(u_0)\Big[\,\bar u_0 \Big(\,\frac{\lambda}{2}
\varphi_\pi(u_0)+\mu_\pi\varphi_{p}(u_0) - \frac{2}{\lambda}g''_1(u_0)\,\Big)
-\frac{2}{\lambda}g_2(u_0) \nnb \\
&&+\,\frac{4}{\lambda}g'_1(u_0)
-\,\frac{2}{\lambda}\int^{u_0}_0\! d \alpha_1 \int^{1-\alpha_1}_{u_0-\alpha_1}
d\alpha_3\,\frac{1}{\alpha_3^2}\,
\varphi_{\parallel}(\alpha_1,1-\alpha_1-\alpha_3,\alpha_3) \,\Big]\,,
\\[3mm]
\frac{1}{\pi}{\rm Im}{E}_b^3(\lambda,\nu)&=&
\frac{f_\pi}{2}\Theta(u_0)\Big[ -\bar u_0 \Big(\,\frac{\lambda}{2}
\varphi_\pi(u_0)+\mu_\pi\varphi_{p}(u_0) - \frac{2}{\lambda}g''_1(u_0)\,\Big)
+\frac{2}{\lambda}g_2(u_0) \nnb \\
&&-\,\frac{4}{\lambda}g'_1(u_0)
+\,\frac{2}{\lambda}\int^{u_0}_0\! d \alpha_1 \int^{1-\alpha_1}_{u_0-\alpha_1}
d\alpha_3\,\frac{1}{\alpha_3^2}\,
\varphi_{\parallel}(\alpha_1,1-\alpha_1-\alpha_3,\alpha_3) \,\Big]\,.
\eeq
Note that one consequence of the equations of motion in Eq.~(\ref{consEM}),
${\cal F}^3_{b} = - {\cal F}^3_{a}$, is explicitly  satisfied by the sum 
rules.  Combining the above sum rules with Eq.~(\ref{Fsix}) yields,
\beq
F_6(v\cdot p)&=&-\frac{f_\pi}{2 F}\,{\rm e}^{2\bar\Lambda/T}\,
v\cdot p \int^{\theta}_0 {\rm d} u\,
\bigg[\,\frac{\mu_\pi}{6}\varphi_\sigma(\bar u)
-\frac{f_{3\pi}}{f_\pi}\int^{\bar u}_0 d \alpha_1 
\int^{1-\alpha_1}_{\bar u-\alpha_1} d\alpha_3\frac{1}{\alpha_3^2} \nnb \\
&&\qquad\qquad\qquad\qquad\qquad\quad
\times\,\varphi_{3\pi}(\alpha_1,1-\alpha_1-\alpha_3,\alpha_3)\,\bigg]\,
{\rm e}^{-2u\,v\cdot p/T} \,.~
\eeq

\subsection{$1/m_Q$ corrections from the effective Lagrangian}

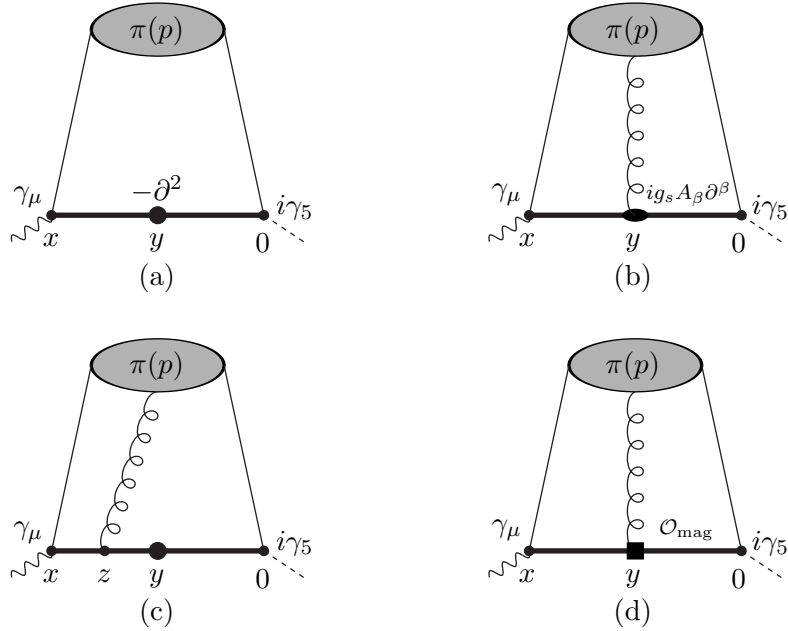
\begin{figure}
\begin{center}
\begin{picture}(320,100)(-5,0)
\Text(70,-3)[t]{(a)}
\GOval(70,85)(10,25)(0){0.7}
\Text(70,85)[]{$\pi (p)$}
\SetWidth{2.0}
\Line(30,15)(110,15)
\SetWidth{0.5}
\Line(30,15)(45,85)
\Line(110,15)(95,85)
\Photon(30,15)(15,5){2}{3}
\DashLine(110,15)(125,5){2}
\Vertex(30,15){2}
\Vertex(110,15){2}
\Text(30,8)[t]{$x$}
\Text(70,8)[t]{$y$}
\Text(110,8)[t]{$0$}
\Vertex(70,15){3.4}
\Text(70,25)[]{$-\partial^2$}
\Text(27,22)[r]{$\gamma_\mu$}
\Text(115,18)[l]{$i\gamma_5$}
\Text(250,-3)[t]{(b)}
\GOval(250,85)(10,25)(0){0.7}
\Text(250,85)[]{$\pi (p)$}
\SetWidth{2.0}
\Line(210,15)(290,15)
\SetWidth{0.5}
\Line(210,15)(225,85)
\Line(290,15)(275,85)
\Photon(210,15)(195,5){2}{3}
\DashLine(290,15)(305,5){2}
\Vertex(210,15){2}
\Vertex(290,15){2}
\GOval(250,15)(2.2,4.4)(0){0.0}
\Text(210,8)[t]{$x$}
\Text(250,8)[t]{$y$}
\Text(290,8)[t]{$0$}
\Text(208,23)[r]{$\gamma_\mu$}
\Text(295,18)[l]{$i\gamma_5$}
\Gluon(250,15)(250,75){3}{5}
\Text(255,23)[l]{${}_{ig_s A_\beta \partial^{\beta} }$}
\end{picture}
\end{center}
\vskip 3mm
\begin{center}
\begin{picture}(320,100)(-5,0)
\Text(70,-3)[t]{(c)}
\GOval(70,85)(10,25)(0){0.7}
\Text(70,85)[]{$\pi (p)$}
\SetWidth{2.0}
\Line(30,15)(110,15)
\SetWidth{0.5}
\Line(30,15)(45,85)
\Line(110,15)(95,85)
\Photon(30,15)(15,5){2}{3}
\DashLine(110,15)(125,5){2}
\Vertex(110,15){2}
\Vertex(70,15){3.4}
\Vertex(50,15){2}
\Vertex(30,15){2}
\Text(70,8)[t]{$y$}
\Text(50,8)[t]{$z$}
\Text(30,8)[t]{$x$}
\Text(110,8)[t]{$0$}
\Text(27,22)[r]{$\gamma_\mu$}
\Text(115,18)[l]{$i\gamma_5$}
\Gluon(50,15)(70,75){3}{6}
\Text(250,-3)[t]{(d)}
\GOval(250,85)(10,25)(0){0.7}
\Text(250,85)[]{$\pi (p)$}
\SetWidth{2.0}
\Line(210,15)(290,15)
\SetWidth{0.5}
\Line(210,15)(225,85)
\Line(290,15)(275,85)
\Photon(210,15)(195,5){2}{3}
\DashLine(290,15)(305,5){2}
\Vertex(210,15){2}
\Vertex(290,15){2}
\Text(210,8)[t]{$x$}
\Text(250,8)[t]{$y$}
\Text(290,8)[t]{$0$}
\Text(208,23)[r]{$\gamma_\mu$}
\Text(295,18)[l]{$i\gamma_5$}
\Gluon(250,15)(250,75){3}{5}
\GBoxc(250,15)(6,6){0.0}
\Text(260,23)[l]{${}_{{\cal O}_{\rm mag}}$}
\end{picture}
\end{center}
\caption{\it The diagrammatic representation of the leading order vacuum-pion
correlation functions with ${\cal O}_{\rm kin}$ and ${\cal O}_{\rm mag}$
insertions, (see Eq.(\ref{EffL})), where the heavy solid lines
represent the heavy quark, and the helical lines a gluon.
\label{Fig_ES}}
\end{figure}

Finally, we consider the $1/m_Q$ corrections to the correlation function 
Eq.~(\ref{CFLO}) from the sub-leading operators ${O}_{\rm kin}$ and 
${O}_{\rm mag}$ in Eq.~(\ref{operKG}).  They are depicted in 
Fig.~\ref{Fig_ES}.  One has
\vspace*{2mm}
\beq
\label{EffL}
F^{(\frac{1}{m_Q})}_\mu(\lambda,\tilde \omega)
&=&i\!\int\! {\rm d}^4 x\,{\rm e}^{ik\cdot x}\,\langle \pi(p)|\,
T\bigg\{ \bar u(x) \gamma_\mu h_v(x),\,
i\!\int\! {\rm d}^4y\, {\cal L}_{1}(y)\,,
\bar h_v(0) i \gamma_5 d(0) \bigg\} \,| 0 \rangle \nnb \\
&=&F^{(\frac{1}{m_Q})}_a (\lambda,\tilde \omega)v_\mu +
F^{(\frac{1}{m_Q})}_b (\lambda,\tilde \omega) \widehat p_\mu \,,
\eeq
where
${\cal L}_1 = {\cal O}_{\rm kin}+C_{\rm mag}{\cal O}_{\rm mag}$.
By denoting $\delta L_\alpha /2m_Q$ as the $1/m_Q$ corrections to $L_\alpha$,
namely
\beq
\delta L_\alpha &\equiv & K_\alpha + C_{\rm mag}\, {\cal S}_\alpha\,,
\eeq
the hadronic representation of this correlator can be written as~\cite{dhll}
\beq
F^{(\frac{1}{m_Q})\,\rm Hadr}_{\alpha}(\lambda, \tilde{\omega})&=&
\frac{2 F\,\delta F\, L_\alpha(\lambda)}{2\bar\Lambda-\tilde\omega}
+\frac{2 F\,\delta L_\alpha(\lambda)}{2\bar\Lambda-\tilde\omega}
-\frac{4 F\,\delta\bar\Lambda \, L_\alpha(\lambda)}
{(2\bar\Lambda-\tilde\omega)^2}\,+\,{resonances}\,,
\eeq
where $\delta F/2m_Q$ and $\delta\bar\Lambda/2m_Q$ are the $1/m_Q$ corrections
for $F$ and $\bar\Lambda$~\cite{N_BB,N_B}, respectively.

We now calculate the sub-leading correlation function Eq.~(\ref{EffL}) in  
HQET, starting from the relation
\beq \label{G_approx}
\partial^\beta A_{\beta}&\cong & {\cal W F}(4){\rm s} +
{\cal O}({\alpha_{\it s}})\,{\cal WF}(2){\rm s}\,,
\eeq
where the symbol $\cong$ indicates that this equation holds at the level of 
matrix elements in terms of LC wave functions. This relation can be found by
considering Eq.(\ref{A_fixed_point}) and the equation of motion for the gluon
fields,
$\partial^{\alpha}G^{a}_{\alpha\beta}\!=\! - f^{abc}A^{b\,\alpha} 
G^{c}_{\alpha\beta}+ g_s \bar h_v \:\!v_\beta\, T^a\:\! h_v$.
One can drop all the terms containing $\partial^\beta A_{\beta}$ or 
$A^\beta A_\beta$, because contributions from  wave functions 
with more than $3$ particles and from order $\alpha_{\it s}$ contributions are 
physically quite small~\cite{lcsr}. Therefore, we have
\beq
F^{(\frac{1}{m_Q})}_{\mu}(\lambda,\tilde \omega)
&=& i\!\int {\rm d}^4 x\,{\rm e}^{ik\cdot x}\,i\!\int\! {\rm d}^4y \;
\langle \pi(p)|\,T\bigg\{ \,\bar u(x) \gamma_\mu h_v(x), \,\bar h_v(y)
\Big[ -\partial^2+2\, i\, g_s A_\beta \partial^{\beta} \nnb \\
&&\qquad\qquad\qquad\qquad\qquad\qquad
+\,C_{\rm mag}\frac{g_{\it s}}{2}\sigma_{\alpha\beta}G^{\alpha\beta}\,\Big]
h_v(y), \, \bar h_v(0) i \gamma_5 d(0)\,\bigg\} \,| 0 \rangle \,. ~~~~~
\eeq
Here the higher-order terms in the heavy quark propagator may be included,
as displayed in Fig.~\ref{Fig_ES} (c). However, direct evaluation shows
their contribution to be zero.
Using the identities Eqs.~(\ref{Id1}) and (\ref{Id2}) together with
\beq
\gamma_\mu\sigma_{\alpha\beta}&=&
i\,(g_{\mu\alpha}\gamma_\beta-g_{\mu\beta}\gamma_\alpha)
+\varepsilon_{\mu\alpha\beta\delta}\gamma^\delta\gamma_5\,, \\[3mm]
\gamma_\mu\sigma_{\alpha\beta}\gamma_\nu&=&
i\,(g_{\mu\alpha}g_{\beta\nu}-g_{\mu\beta}g_{\alpha\nu})
+ (\sigma_{\beta\nu}g_{\mu\alpha}-\sigma_{\alpha\nu}g_{\mu\beta})
-\varepsilon_{\mu\alpha\beta\nu}\gamma_5
+ i\,\varepsilon_{\mu\alpha\beta\delta}g^{\delta\rho}
\sigma_{\rho\nu}\gamma_5\,,
\eeq
and after some algebraic manipulations, we obtain 
\beq
F^{(\frac{1}{m_Q})}_{\mu}(\lambda,\tilde \omega)
\!&=&\!\frac{f_\pi}{2}\int_0^1\!{\rm d} u \int^\infty_0\! {\rm d}t \,
{\rm e}^{i\frac{\tilde\omega}{2}t }\,\Bigg\{\,
{\frac{\scriptstyle\overleftarrow{\rm d^2}}{{\rm d} t^2}}
\bigg[\,\frac{\lambda}{2}\Big[\,t\varphi_\pi(u)+t^3g_1(u)\,\Big]\widehat p_\mu
+\mu_\pi t\varphi_p(u)v_\mu \nnb \\
&& +\,it^2\Big[\,\frac{\mu_\pi}{12}\lambda \varphi_\sigma(u)-g_2(u)\,\Big]
(\widehat p_\mu-v_\mu)\, \bigg]
+{\frac{\scriptstyle\overleftarrow{\rm d}}{{\rm d} t}}
\bigg[\: 2it\Big[\,\frac{\mu_\pi}{12}\lambda \varphi_\sigma(u)
-g_2(u)\,\Big](\widehat p_\mu-v_\mu) \nnb \\
&& +\,2\lambda t^2 g_1(u)\widehat p_\mu -i\bar u \lambda \bigg(\,
\frac{\lambda}{2}\Big[\,t\varphi_\pi(u)+t^3g_1(u)\,\Big]\widehat p_\mu
+\mu_\pi t\varphi_p(u)v_\mu \nnb \\
&& +\,it^2\Big[\,\frac{\mu_\pi}{12}\lambda \varphi_\sigma(u)
-g_2(u)\,\Big](\widehat p_\mu-v_\mu)\,\bigg)\:\bigg] + \bigg(\,
\lambda t g_1(u)+2ig_2(u) \nnb \\
&& -\,i\bar u\lambda
\Big[\,\lambda t^2 g_1(u)-it g_2(u)\,\Big]\,\bigg)\widehat p_\mu\,
\,\Bigg\}\,{\rm e}^{-i\bar u\frac{\lambda}{2}t } \,
- f_\pi\! \int \!{\cal D}\alpha_i\!
\int_0^1\!{\rm d}s\,s \int_0^1\!{\rm d}w\,w
\int^\infty_0\!{\rm d}\, t \,
{\rm e}^{i\frac{\tilde\omega}{2}t}  \nnb \\
&&\times\, \frac{\lambda}{2} t\,\widehat p_\mu\, \bigg\{
\,2\Big[\,\varphi_{\parallel}(\alpha_i)\,
+\varphi_{\perp}(\alpha_i)\,\Big]\,
-i(1-\alpha_1-w\alpha_3)\;\!\frac{\lambda}{2}t\,\varphi_{\parallel}(\alpha_i)
\,\bigg\}\,{\rm e}^{-i(1-\alpha_1-sw\alpha_3)\frac{\lambda}{2}t} \nnb \\
&&+\,C_{\rm mag}\frac{f_\pi}{2}\!\int\! {\cal D}\alpha_i\!
\int^1_0\! {\rm d}w \int_o^\infty\! {\rm d}t\, 
{\rm e}^{i\frac{\tilde\omega}{2}t}\, \frac{\lambda}{2}t\,\bigg\{
-\Big[\,2\widetilde \varphi_{\perp}(\alpha_i)-\tilde \varphi_{\parallel}
(\alpha_i)+ 2\varphi_{\perp}(\alpha_i) \,\Big]\,\widehat p_\mu \nnb \\
&& \qquad\qquad\qquad\qquad\qquad\qquad\qquad
+\,2\varphi_{\perp}(\alpha_i)\,v_\mu\, \bigg\}\,
{\rm e}^{-i(1-\alpha_1-w\alpha_3)\frac{\lambda}{2}t}\,.
\eeq

Having all necessary results at hand, we obtain the sum rules resulting from 
order $1/m_Q$ power corrections to the effective Lagrangian:
\beq
\delta L_\alpha(\lambda)&=&\frac{1}{2F}
\int^{\omega_c}_0 {\rm d}\nu\,\frac{1}{\pi}\,{\rm Im}\,
{F}^{(\frac{1}{m_Q})}_{\alpha}(\lambda,\nu)
\,{\rm e}^{(2\bar\Lambda-\nu)/T}
\,+\Big[{2\,\delta\bar\Lambda/T-\delta F}\Big]
{L}_{\alpha}(\lambda)\,,
\eeq
where the sub-leading HQET spectral density functions are given by
\beq
\frac{1}{\pi}{\rm Im}F^{(\frac{1}{m_Q})}_a(\lambda,\nu)&=&
\frac{f_\pi}{2}\Theta(u_0)\bigg\{\,
\bar u_0^2 \Big[\,\mu_\pi\varphi_{p}'(u_0)+\frac{\mu_\pi}{6}\varphi''_{\sigma}
(u_0)-\frac{2}{\lambda} g''_2(u_0)\,\Big]-2\bar u_0\Big[\,\mu_\pi\varphi_{p}
(u_0) \nnb \\
&&\qquad\qquad\quad
+\,\frac{\mu_\pi}{6}\varphi'_{\sigma}(u_0)-\frac{2}{\lambda} g'_2(u_0)\,\Big]
\:\bigg\} \,,
\\[3mm]
\frac{1}{\pi}{\rm Im}F^{(\frac{1}{m_Q})}_b(\lambda,\nu)&=&
\frac{f_\pi}{2}\Theta(u_0)\Bigg\{\,\bar u_0^2\Big[\,
\frac{\lambda}{2}\varphi'_\pi(u_0)-\frac{\mu_\pi}{6}\varphi''_{\sigma}(u_0)
+\frac{2}{\lambda}g_2''(u_0) -\frac{2}{\lambda}g_1'''(u_0)\,\Big] \nnb \\
&&-\,2\bar u_0\Big[\,\frac{\lambda}{2}\varphi_\pi(u_0)-
\frac{\mu_\pi}{6}\varphi'_{\sigma}(u_0)
+\frac{2}{\lambda}g_2'(u_0) -\frac{6}{\lambda}g_1''(u_0)\,\Big]
+\frac{8}{\lambda}g_2(u_0) -\frac{12}{\lambda}g_1'(u_0)  \nnb \\
&&+\,\frac{8}{\lambda}\int^{u_0}_0\!{\rm d}\alpha_1
\int^{1-\alpha_1}_{u_0-\alpha_1}\!{\rm d}\alpha_3\,\frac{1}{\alpha_3^2}\,
\bigg[\,\Big(\frac{\alpha_3}{u_0-\alpha_1}-1\Big)\,
\varphi_{\parallel}(\alpha_1,1-\alpha_1-\alpha_3,\alpha_3)  \nnb \\
&&-\,\ln\Big(\frac{\alpha_3}{u_0-\alpha_1}\Big)
\Big[\,\varphi_{\parallel}+\varphi_{\perp}\,\Big]
(\alpha_1,1-\alpha_1-\alpha_3,\alpha_3)\,\bigg]\:\Bigg\}\,.
\eeq
It is interesting to note that the operator ${\cal O}_{mag}$ in fact does not  
contribute to the HQET spectral density functions even at order $1/m_Q$.

\section{Numerical analysis} \label{Numerical}

Let us now analyze the sum rules numerically.  We use the pion wave
functions collected and carefully discussed in Ref.~\cite{BF,lcsr}. 
They read,
\beq
\varphi_\pi(u,\mu)&=&\displaystyle
6u\bar{u}\left\{1+a_2(\mu)\frac{3}{2}[5(u-\bar{u})^2-1]+
a_4(\mu)\frac{15}{8}[21(u-\bar{u})^4-14(u-\bar{u})^2+1]\right\}\,, \nnb \\[1mm]
\varphi_p(u,\mu)  &=&\displaystyle
1+\tilde B_2(\mu)\frac{1}{2}[3(u-\bar{u})^2-1]+
\tilde B_4(\mu)\frac{1}{8}[35(u-\bar{u})^4 -30(u-\bar{u})^2+3]\,, \nnb \\[1mm]
\varphi_\sigma (u,\mu)&=&\displaystyle
6u\bar{u}\bigg\{1+\tilde C_2(\mu)\frac{3}{2}[5(u-\bar{u})^2-1]+
\tilde C_4(\mu)\frac{15}{8}[21(u-\bar{u})^4-14(u-\bar{u})^2+1]\bigg\}\,, 
\nnb \\[1mm]
g_1(u,\mu)&=&\displaystyle
\frac{5}{2}\delta^2(\mu)\bar{u}^2u^2+\frac{1}{2}\varepsilon(\mu)
\delta^2(\mu)[\bar{u}u(2+13\bar{u}u)+10u^3\ln u(2-3u+\frac{6}{5}u^2) 
\nnb \\[0mm]
&&\displaystyle
+10\bar{u}^3\ln \bar{u}(2-3\bar{u}+\frac{6}{5}\bar{u}^2)]\,, \nnb \\[1mm]
g_2(u,\mu)&=&\displaystyle \frac{10}{3}\delta^2(\mu)\bar{u}u(u-\bar{u})\,, 
\nnb \\[0mm]
\varphi_{3\pi}(\alpha_i,\mu)&=&\displaystyle 360 \alpha_1\alpha_2\alpha_3^2
\Big[\,1+\omega_{1,0}(\mu)\frac12(7\alpha_3-3) +\omega_{2,0}(\mu)
(2-4\alpha_1\alpha_2-8\alpha_3+8\alpha_3^2) \nnb \\[0mm]
&& \displaystyle +\omega_{1,1}(\mu)(3\alpha_1\alpha_2-2\alpha_3+3\alpha_3^2)\,
\Big]\,, \nnb \\[1mm]
\varphi_\perp (\alpha_i,\mu)&=&\displaystyle 30\delta^2(\mu)
(\alpha_1-\alpha_2)\alpha_3^2[\frac13 +2
\varepsilon(\mu) (1-2\alpha_3)]\,, \nnb \\[1mm]
\varphi_\parallel (\alpha_i,\mu)&=&120\displaystyle \delta^2(\mu)
\varepsilon(\mu)(\alpha_1-\alpha_2)\alpha_1\alpha_2\alpha_3\,, \nnb \\[1mm]
\tilde{\varphi}_\perp (\alpha_i,\mu)&=&\displaystyle
30\delta^2(\mu)\alpha_3^2(1-\alpha_3)[\frac13+2
\varepsilon(\mu) (1-2\alpha_3)]\,,  \nnb \\[1mm]
\tilde{\varphi}_\parallel (\alpha_i,\mu)&=&\displaystyle
-120\delta^2\alpha_1\alpha_2\alpha_3[\frac13+
\varepsilon(\mu) (1-3\alpha_3)]\,,
\label{wf10}
\eeq
where $\mu$ is the renormalization scale.  The scale dependence of these
wave functions is given by perturbative QCD.

The light-cone sum rules obtained in HQET, and the pion wave functions
in Eq. (\ref{wf10}) depend on the subtraction point $\mu$.
However, in the lowest order of $\alpha_s$, the value of the scale $\mu$ is
ambiguous. Within HQET, a reasonable choice is
$\mu \simeq 2\bar\Lambda$. Thus we set the renormalization point 
at $\mu_0=1$ GeV.
In this case, $\mu_\pi(\mu_0)=1.65$ GeV, and $r(\mu_0)=1.08$.
The coefficients appearing in the above equation are then taken as~\cite{lcsr}
\beq
a_2(\mu_0)=0.44\,, ~~~~a_4(\mu_0)=0.25\,, ~~~~\tilde B_2(\mu_0)=0.48\,,
~~~~\tilde B_4(\mu_0)=1.15\,, \nnb \\
\tilde C_2(\mu_0)=0.10\,, ~~~\tilde C_4(\mu_0)=0.067\,,
~~~\delta^2(\mu_0)=0.2\,{\rm GeV}^2\,, ~~~\varepsilon(\mu_0)=0.5\,,
\nnb \\
f_{3\pi}(\mu_0)=0.0035\, {\rm GeV}^2\,,~~\omega_{1,0}(\mu_0)=-2.88\,,~~
\omega_{2,0}(\mu_0)=10.5\,,~~\omega_{1,1}(\mu_0)=0\,.~
\eeq
The pion decay constant $f_\pi=0.132$ GeV is taken from experiment. 
The parameters for heavy mesons are obtained from HQET sum
rules~\cite{N_BB,N_B,MNrep}:
$\bar\Lambda=0.57$ GeV, $F(1{\rm GeV})=0.46$ GeV$^{3/2}$,
$\delta\bar\Lambda =-0.35$ GeV$^2$ and $\delta F= - 1.92$ GeV.

\subsection{Universal functions}

\begin{figure}[htb]
\vspace*{-0.0cm}
\centerline{
   \epsfysize=7cm \epsfxsize=11cm
   \epsfbox{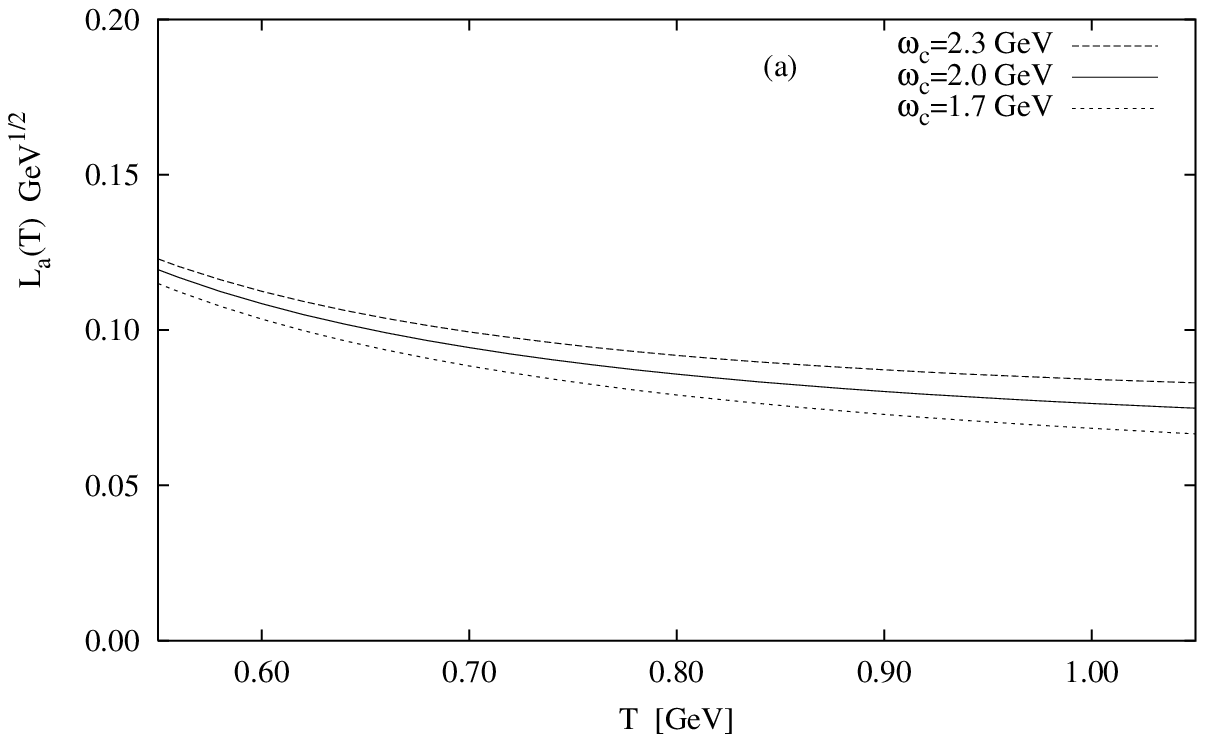} 
}   
   \vspace*{3mm}
\centerline{
   \epsfysize=7cm \epsfxsize=11cm
   \epsfbox{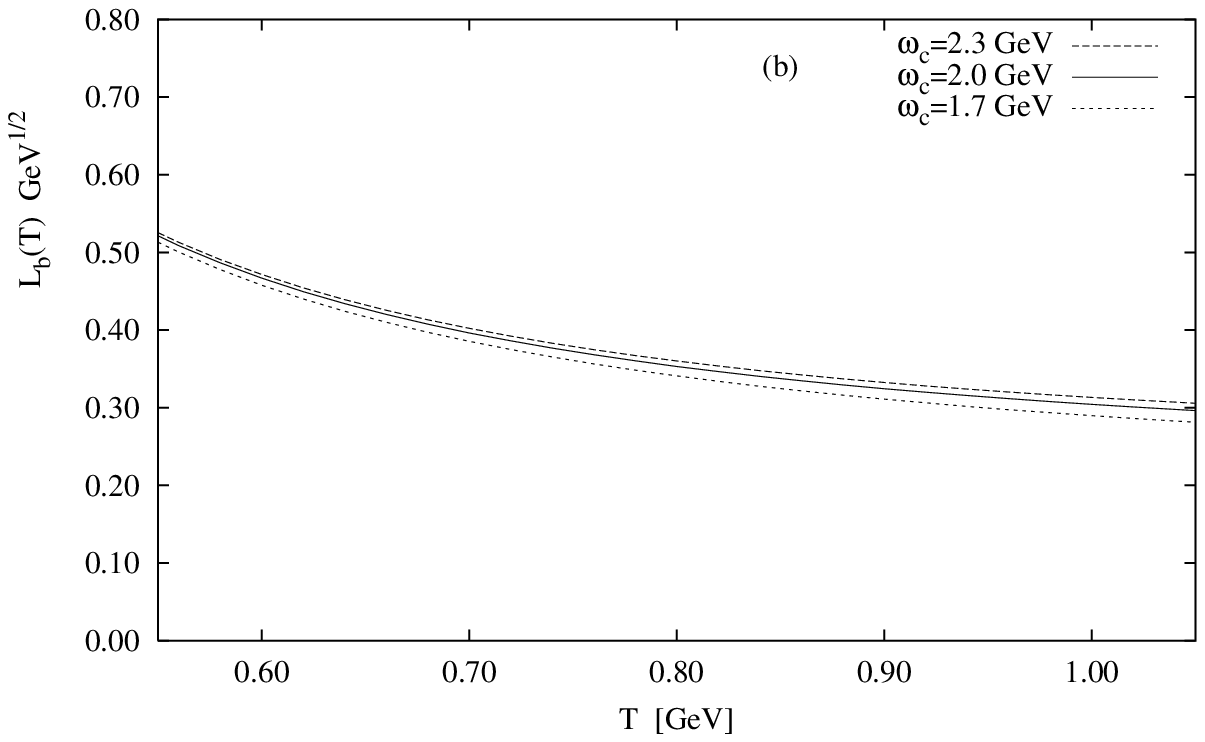}
}
\vspace*{-1mm}
\caption{\it Variation of the sum rule predictions on $L_{a,\,b}$ with the 
Borel parameter $T$ at various values of the continuum threshold parameter 
$\omega_c$.
\label{Win_LO}}
\end{figure}

\begin{figure}[htb]
\vspace*{-0.0cm}
\centerline{
   \epsfysize=7cm \epsfxsize=11cm
   \epsfbox{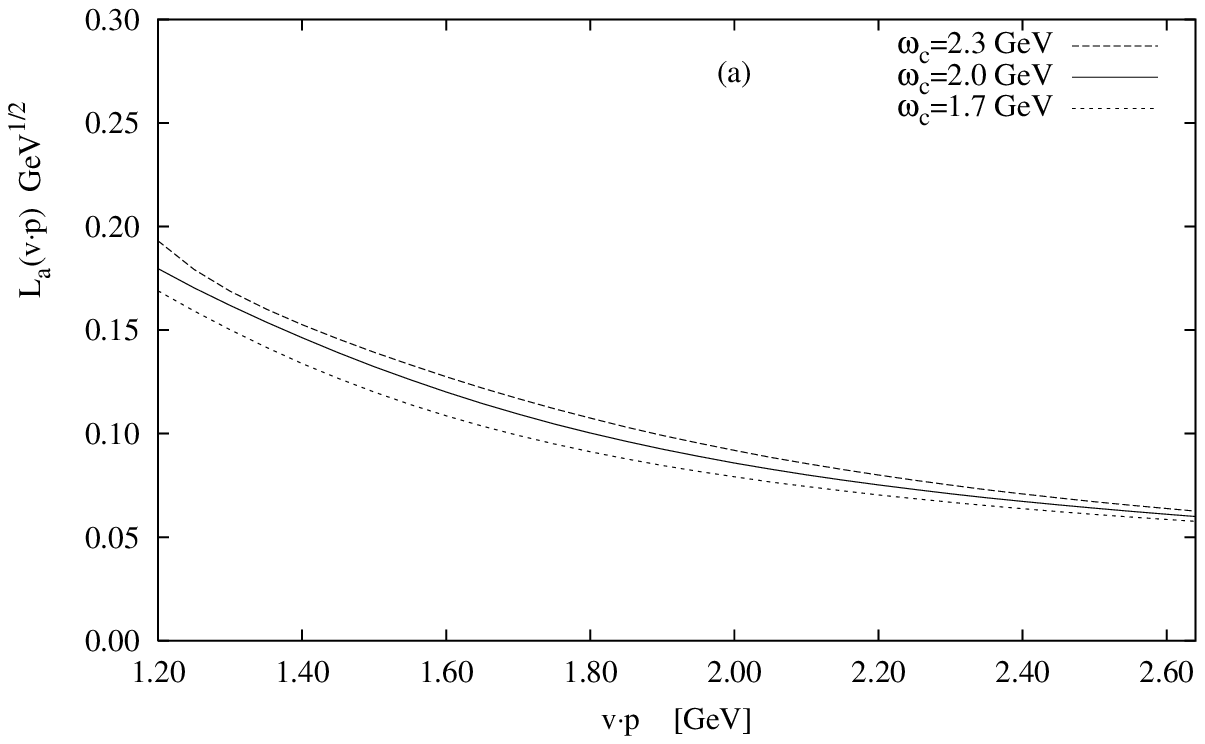} 
}
  \vspace*{3mm}
\centerline{
   \epsfysize=7cm \epsfxsize=11cm
   \epsfbox{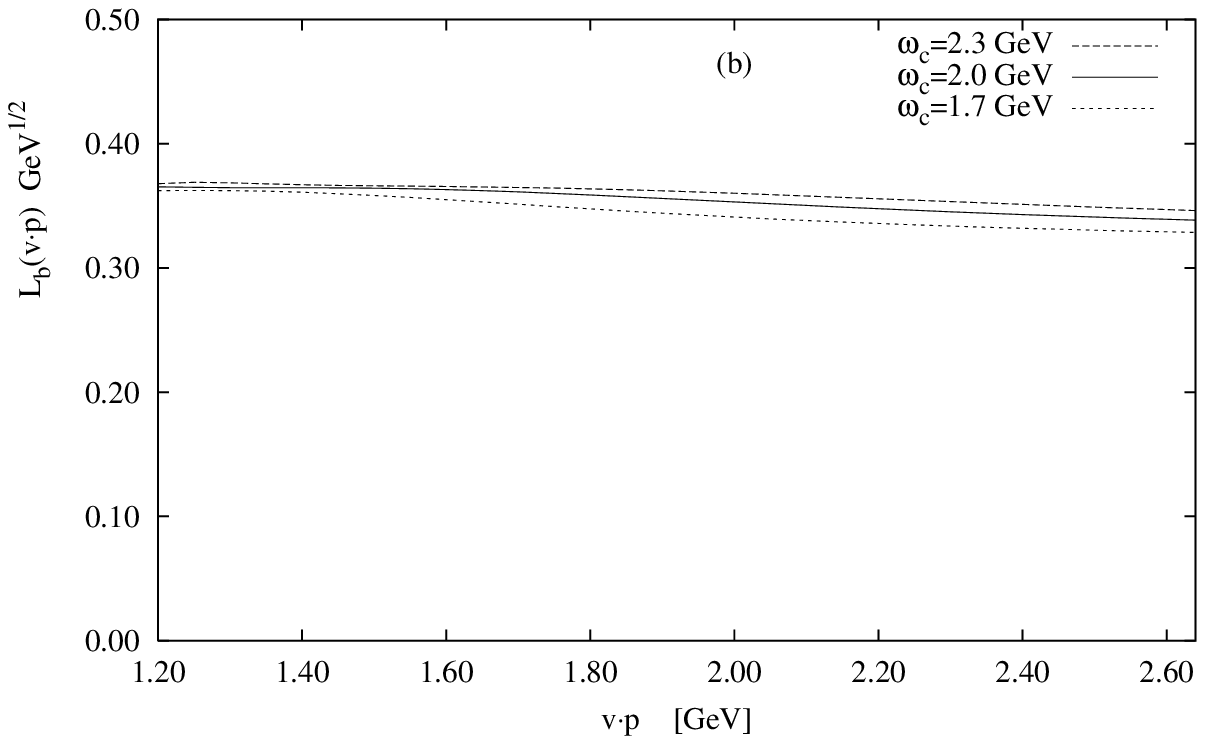} 
}
\vspace*{-1mm}
\caption{\it The universal functions $L_{a,\,b}(v\cdot p)$ obtained from 
LC-HQET sum rules at various values of the continuum threshold parameter 
$\omega_c$.
\label{SR_LO}}
\end{figure}

In order to make the sum rules meaningful, we must first check the existence 
of sum rule windows, which are roughly given by 
$\Lambda_{\rm QCD}<T<2\bar\Lambda$.  The lower and upper limit will be 
obtained by the physical requirement that the Borel parameter $T$ must be 
large enough to ensure that the higher-twist wave function contributions are 
suppressed, and at the same time small enough in order to make the resonance 
contribution not too large.
To be specific, for the typical threshold values $\omega_c \simeq 2$ GeV, we 
find that, setting the Borel parameter to $0.60$ GeV $\leq T\leq 1.00$ GeV 
for the sum rules $L_a$ and $L_b$, the twist-4 wave functions give 
contributions less than $17\%$ and $3\%$, while the resonance contributions 
are lower than $27\%$ and $6\%$, respectively.  The two sum rules are quite 
stable in this region, as shown in Fig.~\ref{Win_LO}(a) and (b) where 
$v\cdot p$ is fixed at $2.0$ GeV.
Nevertheless, our numerical calculations show that the stability of the sum
rules is not very sensitive to a change of $\omega_c$. In order to be 
consistent with HQET sum rule calculations, we adopt the 
same stability region of $\omega_c$ as that from the $\bar\Lambda$ sum 
rule~\cite{N_BB,MNrep}, {\it i.$\,$e.}, $\omega_c=2.0\pm 0.3$ GeV.
Fig.~\ref{SR_LO}(a) and (b) present the final results for $L_a(v\cdot p)$ and
$L_b(v\cdot p)$, respectively, where the central value of the Borel parameter 
$T=0.80$ GeV is used.
Note that the LC-HQET sum rules are meaningless in the soft pion region
as mentioned above. To stay away from the soft pion region, 
we take $v\cdot p > 1.2$ GeV.

For the convenience of further applications, we parametrize the results by 
the following formulae in the region $1.2~{\rm GeV} < v \cdot p <2.64$ GeV,
\beq
L_{a\,}^{\rm Fit}(v\cdot p)&=& \frac{1}{a_0+a_1\,v\cdot p+a_2\,(v\cdot p)^2}
\,,  \\
L_{b\,}^{\rm Fit}(v\cdot p)&=& b_0+b_1\,v\cdot p +\frac{ b_2 }{~ v\cdot p ~}\,.
\label{Pole}
\eeq
Best fit values in which the maximal errors are less than $2\%$ yields the
following values of the parameters:
\begin{center}
\begin{tabular}{c|cccccc}
\hline
$\omega_c~{\scriptstyle {\rm [GeV]}}$~~ & $a_0~{\scriptstyle 
[{\rm GeV}^{-1/2}]}$~~
& $a_1~{\scriptstyle [{\rm GeV}^{-3/2}]}$~~ & $a_2~{\scriptstyle 
[{\rm GeV}^{-5/2}]}$~~~~
& $b_0~{\scriptstyle [{\rm GeV}^{1/2}]}$~~~ & $b_1~{\scriptstyle 
[{\rm GeV}^{-1/2}]}$~~~
& $b_2~{\scriptstyle [{\rm GeV}^{3/2}]}$~~~   \\
\hline
2.3  & -0.78 & 4.00  & 0.891  &  0.403  & -0.0221 & -0.0385 \\
2.0  & -3.03 & 6.71  & 0.286  &  0.409  & -0.0263 & -0.0448 \\
1.7  & -6.25 & 10.9  & -0.743 &  0.320  & -0.0049 & 0.0343  \\
\hline
\end{tabular}
\end{center}

\begin{figure}[htb]
\vspace*{-0.0cm}
\centerline{
   \epsfysize=7cm \epsfxsize=11cm
   \epsfbox{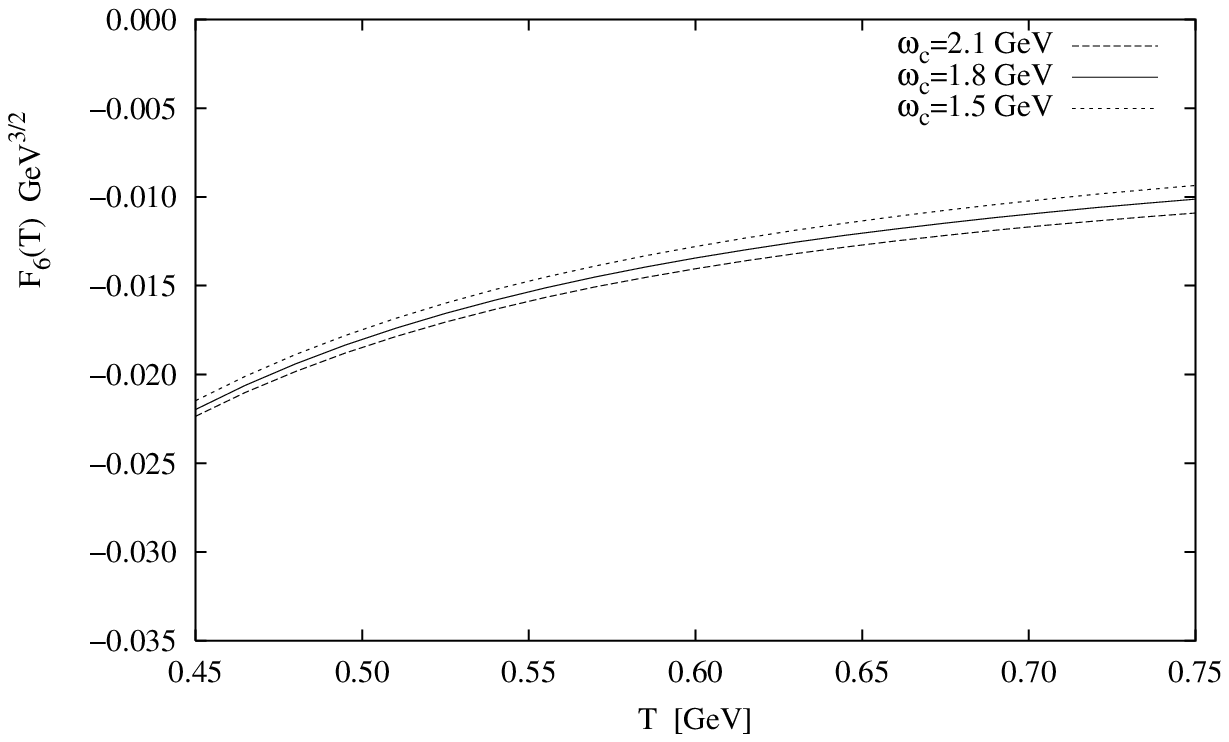} 
}
\vspace*{-3mm}
\caption{\it Dependence of the sum rule predictions of $F_6$ on the Borel
parameter $T$ at various values of the continuum threshold parameter 
$\omega_c$.
\label{Win_F6}}
\end{figure}

\begin{figure}[htb]
\vspace*{0.0cm}
\centerline{
   \epsfysize=7cm \epsfxsize=11cm
   \epsfbox{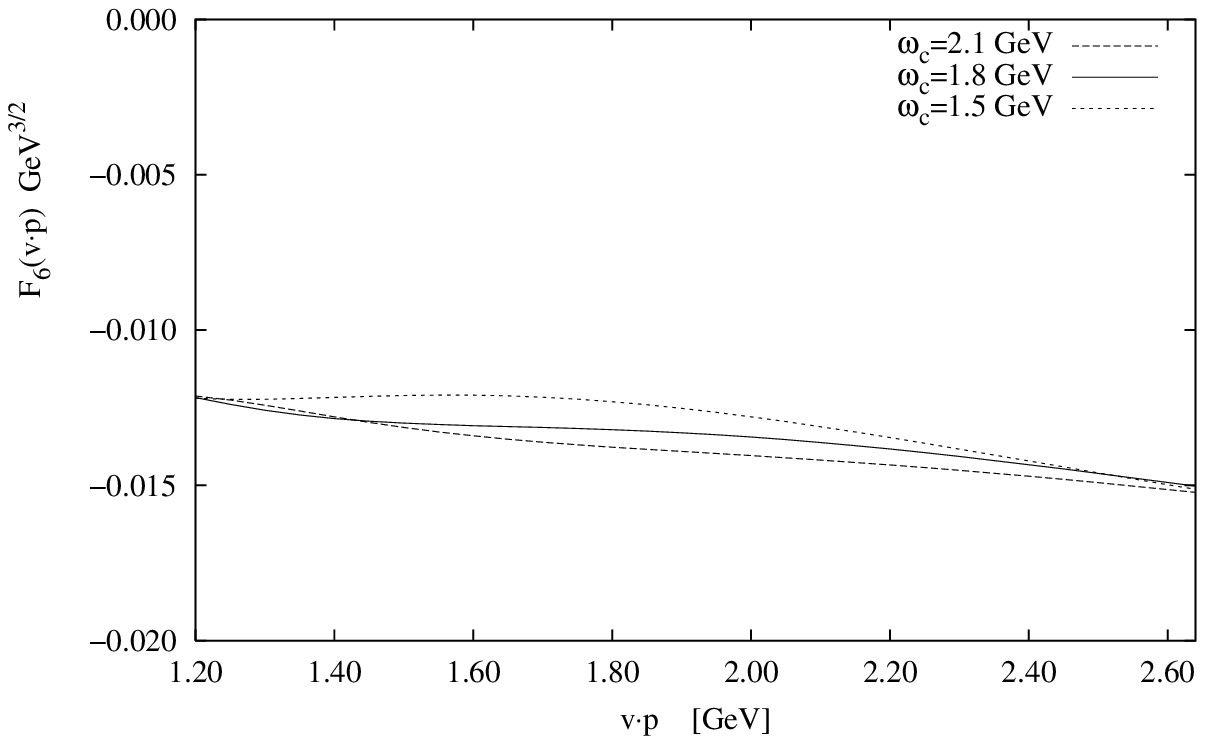}
}
\vspace*{-1mm}
\caption{\it The sub-leading universal function $F_6(v\cdot p)$ obtained from 
the LC-HQET sum rules at various values of the continuum threshold parameter 
$\omega_c$.  
\label{SR_F6}}
\end{figure}

Next we consider the $1/m_Q$ corrections from the effective current.  In the 
relevant universal function $F_6$, only two twist-3 wave functions appear.  
Thus, in order to determine the sum rule window, especially its lower limit, 
one may consult 
the sum rules for ${\cal F}^1_{a,\,b}$ that will contribute to the form
factors 
directly while the renormalization-group effects are not considered. For 
threshold values of $\omega_c \simeq 1.8$ GeV, requiring that both the twist-4 
wave functions and the resonance contributions do not exceed $40\%$, the sum 
rule window can be determined to be $0.50$ GeV $\leq T\leq 0.70$ GeV, in which 
we find that the resonance contributions for $F_6$ are less than $8\%$. The 
numerical results for the universal function $F_6$ as a function of the Borel 
parameter $T$ are shown in Fig.~\ref{Win_F6}, where $v\cdot p$ is fixed to be 
$2.0$ GeV.
Fig.~\ref{SR_F6} presents the results for $F_6(v\cdot p)$, where the central 
value $T=0.60$ GeV is taken for the Borel parameter.

The results can be parametrized by the formula below valid in the region
$1.2~{\rm GeV}< v \cdot p <2.64$ GeV, with maximal errors less than $3\%$,
\beq
F_{6\,}^{\rm Fit}(v\cdot p)&=& \frac{1}{c_0+c_1\,v\cdot p+c_2\,(v\cdot p)^2}\,.
\label{Fit_F6}
\eeq
The parameters $c_i$ are given by 
\begin{center}
\begin{tabular}{c|cccccc}
\hline
$\omega_c~{\scriptstyle {\rm [GeV]}}$~~~~~
& $c_0~{\scriptstyle [{\rm GeV}^{-3/2}]}$~~~~~
& $c_1~{\scriptstyle [{\rm GeV}^{-5/2}]}$~~~~~
& $c_2~{\scriptstyle [{\rm GeV}^{-7/2}]}$~~~~~  \\
\hline
2.1  & -112 & 30.5  & -4.93   \\
1.8  & -86.1 & 1.54  & 2.22   \\
1.5  & -57.6 & -34.0 & 11.7  \\
\hline
\end{tabular}
\end{center}

\begin{figure}[htb]
\vspace*{-0.0cm}
\centerline{
   \epsfysize=7cm \epsfxsize=11cm
   \epsfbox{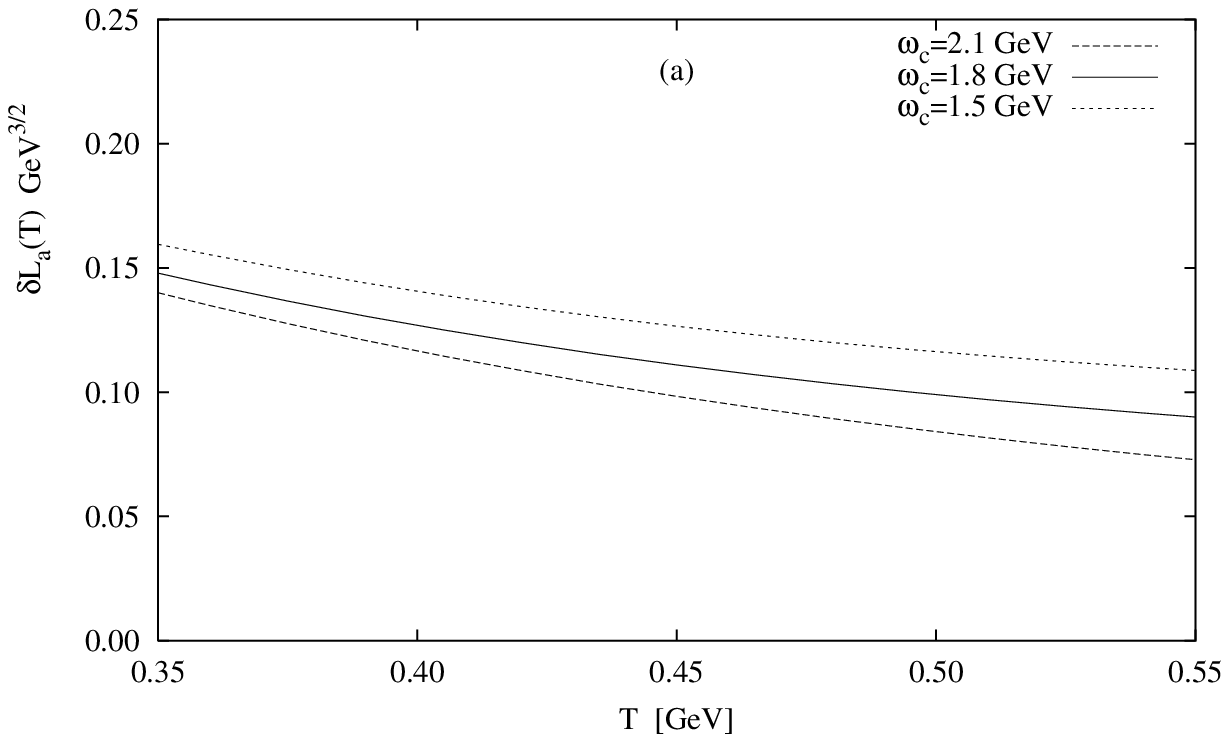} 
}
   \vspace*{3mm}
\centerline{ 
   \epsfysize=7cm \epsfxsize=11cm
   \epsfbox{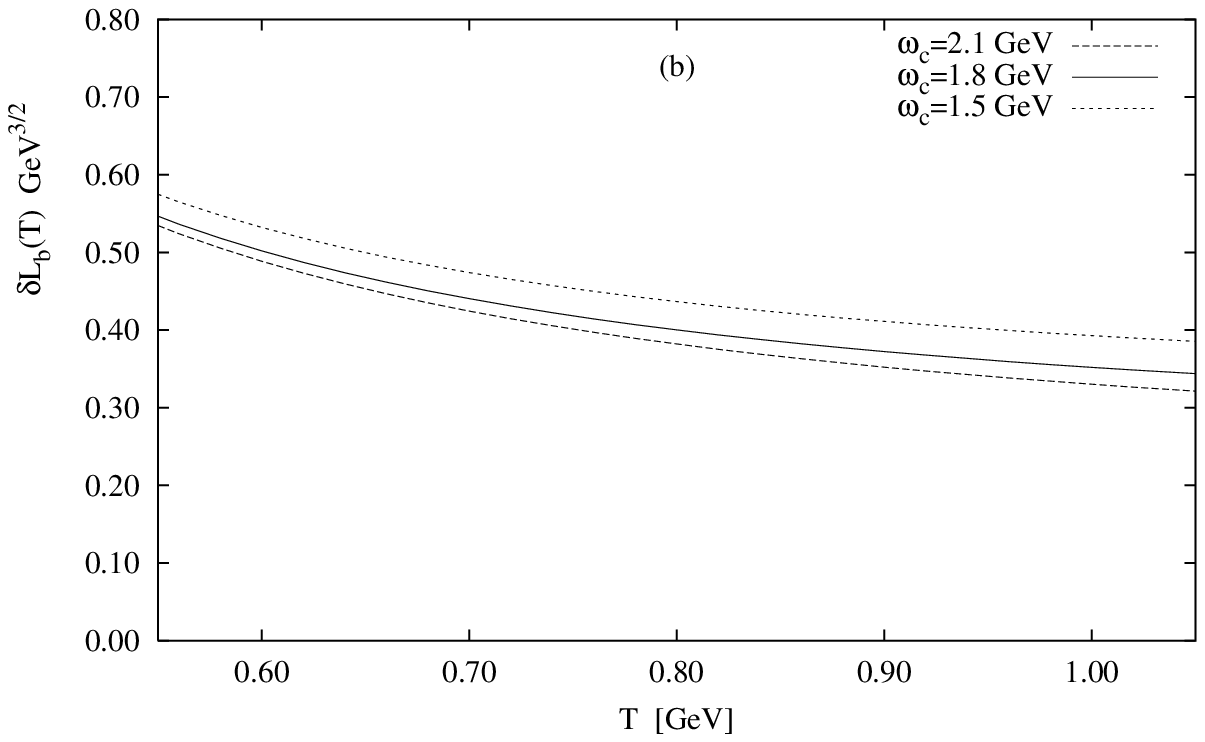}
}
\vspace*{-1mm}
\caption{\it Dependence of the sum rule predictions $\delta L_{a,\,b}$ on the 
Borel parameter $T$ at various values of the continuum threshold parameter 
$\omega_c$.
\label{Win_ES}}
\end{figure}

\begin{figure}[htb]
\vspace*{-0.0cm}
\centerline{
   \epsfysize=7cm \epsfxsize=11cm
   \epsfbox{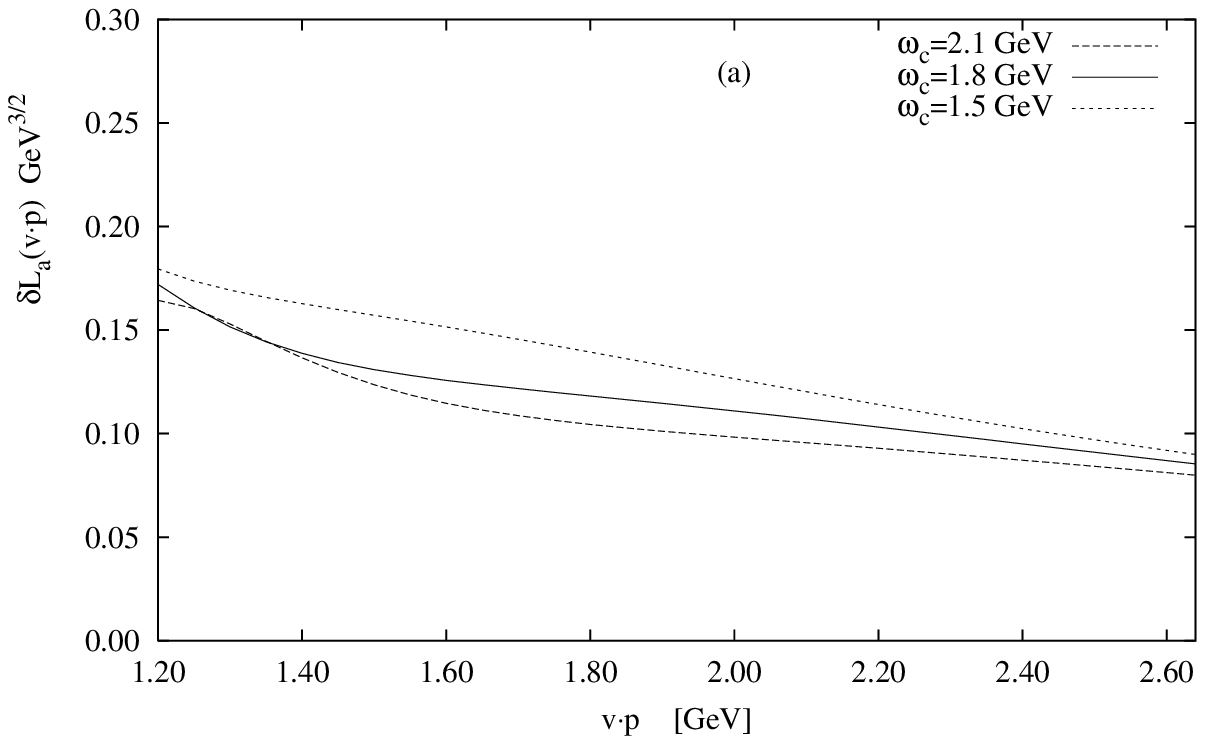} 
}
   \vspace*{3mm}
\centerline{
   \epsfysize=7cm \epsfxsize=11cm
   \epsfbox{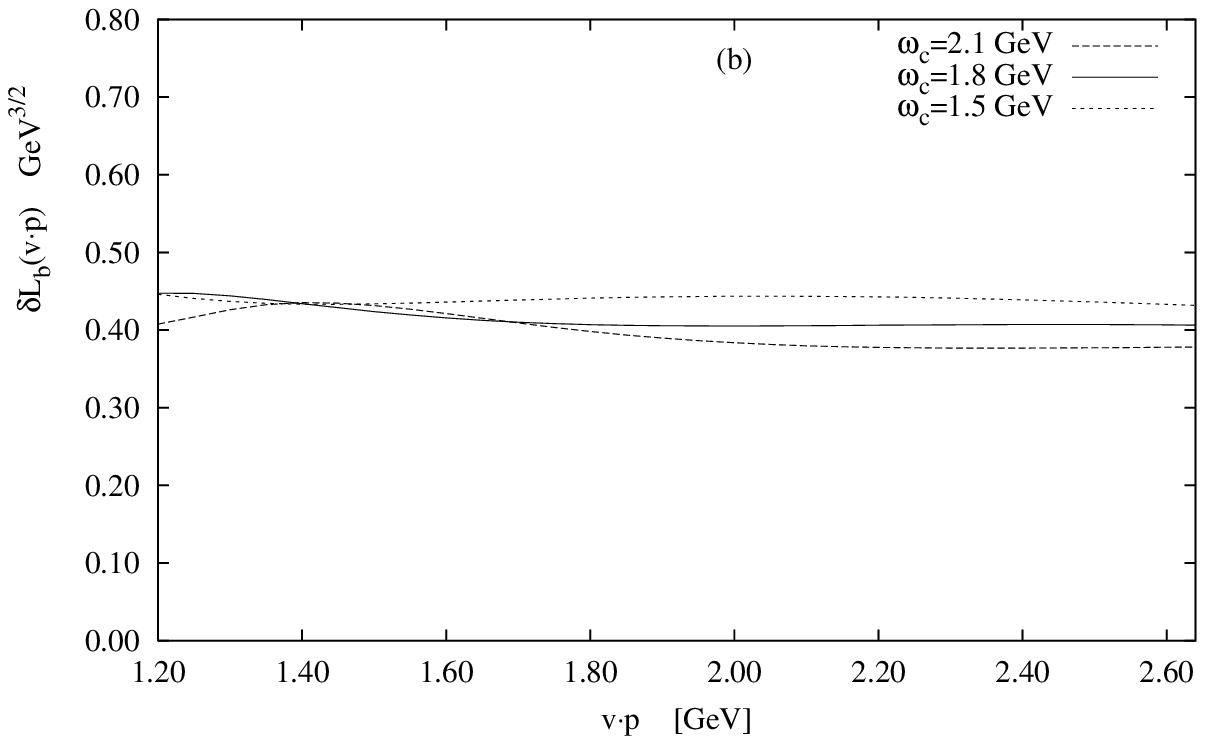}
}
\vspace*{-1mm}
\caption{\it The sub-leading universal functions $\delta L_{a,\,b}(v\cdot p)$
obtained from LC-HQET sum rules at various values of the continuum threshold
parameter $\omega_c$.
\label{SR_ES}}
\end{figure}

We next analyze the sum rules resulting from the $1/m_Q$ insertions due to the
power corrections to the effective Lagrangian.  We find that, for the sum rule 
for $\delta L_a$, the twist-4 wave function contributions are very small 
everywhere in the physically appropiate region of $T$, but the resonance 
contributions grow rapidly as $T$ becomes large. Requiring the latter 
contributions to be less than $40\%$ yields $T\leq 0.55$ GeV.  Consequently, 
we set the range of the Borel parameter at $0.35$ GeV $\leq T\leq 0.55$ GeV.
The results of the sum rule for $\delta L_b$ are very similar to the leading 
order sum rules. In the sum rule window $0.60$ GeV $\leq T\leq 1.00$ GeV and 
for threshold values $\omega_c \simeq 1.8$ GeV, the twist-4 wave function
contributions are less than $11\%$ and the resonance contribution does not 
exceed $18\%$.  The stability of the sum rules for $\delta L_\alpha $ 
($\alpha =a,\,b$) with regard to variations of the Borel parameter $T$ are
shown in 
Fig.~\ref{Win_ES}(a) and (b), where $v\cdot p$ is fixed at $2.0$ GeV.
Fig.~\ref{SR_ES}(a) and (b) present our final results for 
$\delta L_a(v\cdot p)$ and $\delta L_b(v\cdot p)$, respectively, where the
corresponding central value of the Borel parameter $T=0.45$ and $0.80$ GeV are 
used.
It should be mentioned that, in these two sub-leading universal functions, the 
sub-leading HQET spectral density functions give small and negative 
contributions to the form factors, while the corrections coming from 
the $\delta F$ and $\delta\bar\Lambda$ contributions are large, and positive
when the sum is taken 
(for $\delta L_a$, the proporation is about $-1\, : \,2$, and $\delta L_b$ 
about $-1\, : \,5$ ).

We have parametrized these results by the following formulae in the region 
$1.2~{\rm GeV} < v \cdot p <2.64$ GeV,
\beq
\delta L_{a\,}^{\rm Fit}(v\cdot p)&=& \frac{1}{\tilde a_0+\tilde a_1\,v\cdot p
+\tilde a_2\,(v\cdot p)^2}\,,  \\
\delta L_{b\,}^{\rm Fit}(v\cdot p)&=& \tilde b_0+\tilde b_1\,v\cdot p +
\frac{\tilde b_2 }{~ v\cdot p ~}\,.
\label{sub_Pole}
\eeq
The best fitting parameters are
\begin{center}
\begin{tabular}{c|cccccc}
\hline
$\omega_c~{\scriptstyle {\rm [GeV]}}$~~ & $\tilde a_0~{\scriptstyle 
[{\rm GeV}^{-3/2}]}$~~
& $\tilde a_1~{\scriptstyle [{\rm GeV}^{-5/2}]}$~~ & $\tilde a_2~{\scriptstyle 
[{\rm GeV}^{-7/2}]}$~~~~
& $\tilde b_0~{\scriptstyle [{\rm GeV}^{3/2}]}$~~~  & $\tilde b_1~
{\scriptstyle[{\rm GeV}^{1/2}]}$~~~
& $\tilde b_2~{\scriptstyle [{\rm GeV}^{5/2}]}$~~~   \\
\hline
2.1  & -4.56 & 10.6  & -1.59  &  0.319  & 0.0059 & 0.113    \\
1.8  & 1.62  & 3.57  & 0.088  &  0.129   & 0.0623 & 3.00    \\
1.5  & 6.05  & -2.16 &  1.55  &  0.539   & -0.0294 & -0.077 \\
\hline
\end{tabular}
\end{center}
where the maximal fitting errors are less than $5\%$.

Finally we remark on the errors in the sum rules. The uncertainties of the 
continuum threshold parameter $\omega_c$ induce errors (from both the
LC-HQET sum rules themselves and the sum rules of $\bar{\Lambda}$ and
$F$~\cite{MNrep}) less than $\pm\,10\%$ in the final sum rule results.
However, one should keep in mind that the sum rule method typically has 
a $\pm\,(10\sim 30)\%$ uncertainty resulting from the duality assumption,
uncertainties in the wave functions and other input parameters.

\subsection{Form factors to order $1/m_Q$}

\begin{figure}[htb]
\vspace*{-0.0cm}
\centerline{
   \epsfysize=7.7cm \epsfxsize=12cm
   \epsfbox{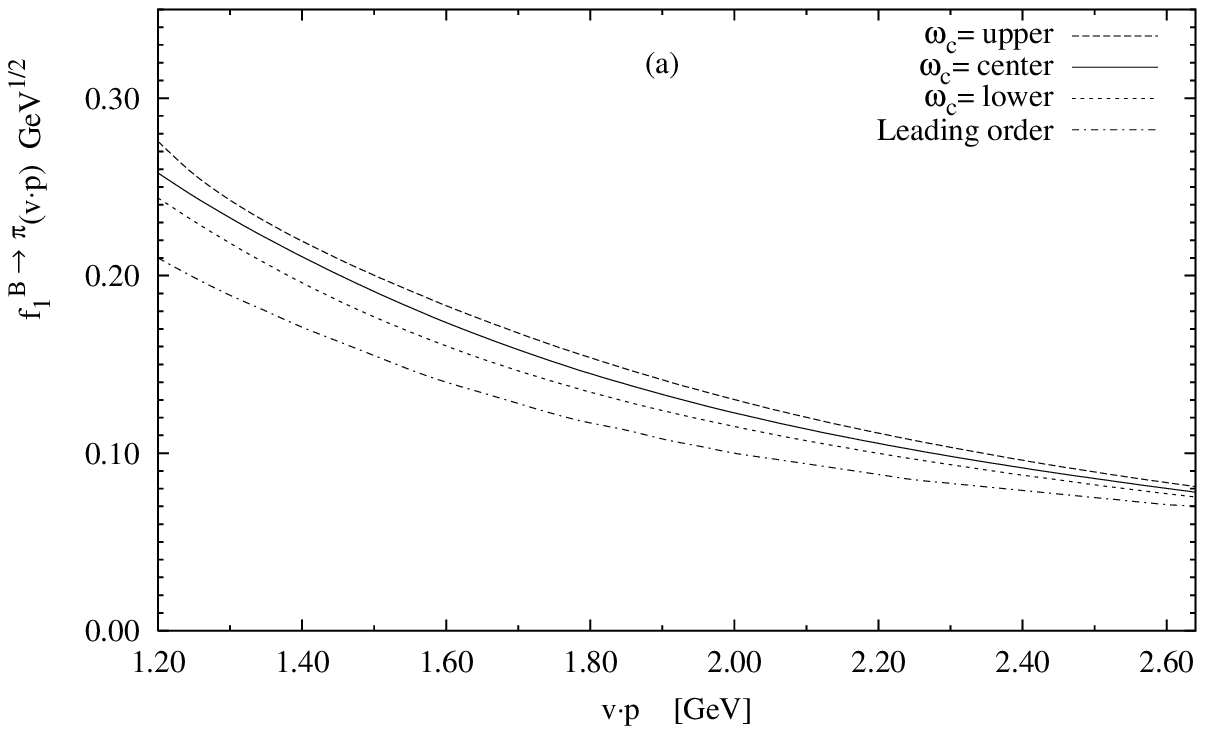} 
}
  \vspace*{3mm}
\centerline{
   \epsfysize=7.7cm \epsfxsize=12cm
   \epsfbox{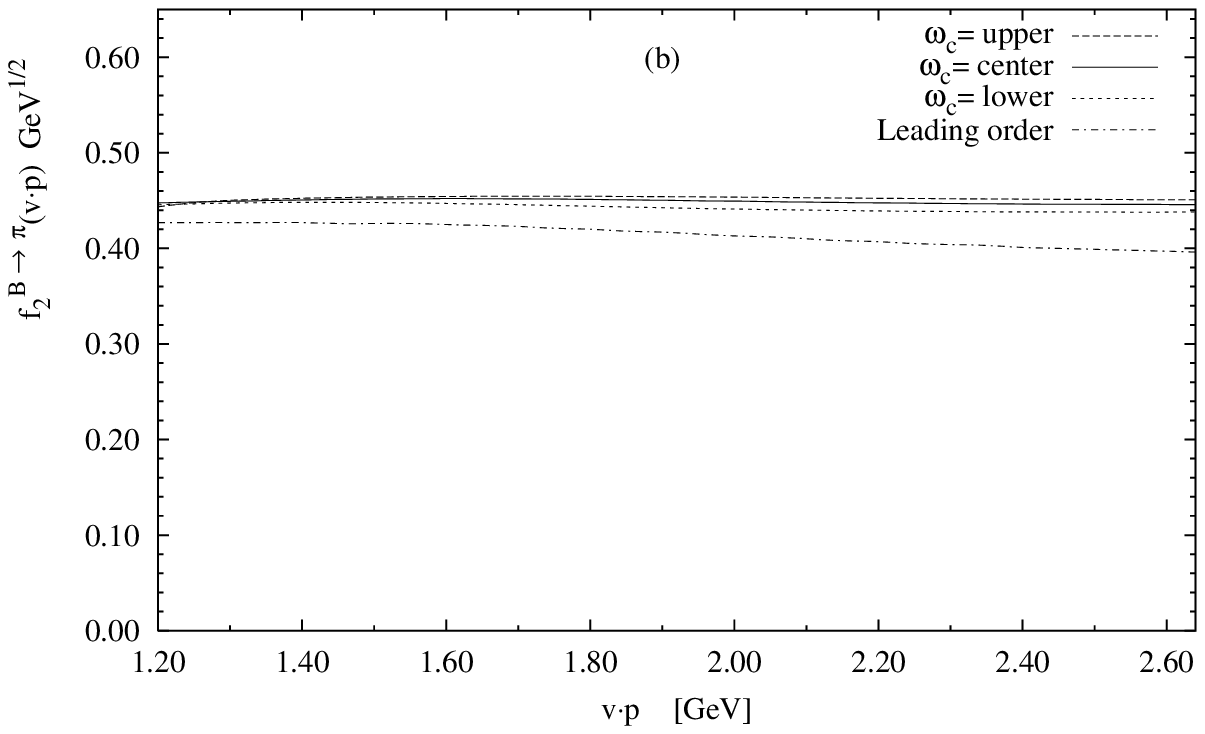}
}
\vspace*{-1mm}
\caption{\it The form factors $f^{B\to \pi}_{1,\,2}(v\cdot p)$ obtained from 
the LC-HQET sum rules at various values of the continuum threshold parameter 
$\omega_c$.  
\label{f12}}
\end{figure}

Having obtained the leading and the relevant sub-leading universal functions,
we can construct the form factors $f_{1,\,2}(v\cdot p)$ from 
Eq.~(\ref{FormFactors}) for the $B\to\pi$ transition to order $1/m_b$. The 
results are shown in Fig.~\ref{f12} (a) and (b), where we take $m_b=4.7$ GeV. 
For comparison, the leading order results corresponding to $\omega_c=2.0$ GeV 
are also shown in this figure. Here the sub-leading universal functions 
roughly give a $11\sim 23\%$ and $5\sim 12\%$ enhancement for the form factors 
$f^{B\to \pi}_{1}(v\cdot p)$ and $f_{2}^{B\to \pi}(v\cdot p)$, respectively.

\section{Comparison with other model calculations, Discussion and Summary}  
\label{Summary}

\begin{figure}[htb]
\vspace*{-0.0cm}
\centerline{
   \epsfysize=7.5cm \epsfxsize=12.5cm
   \epsfbox{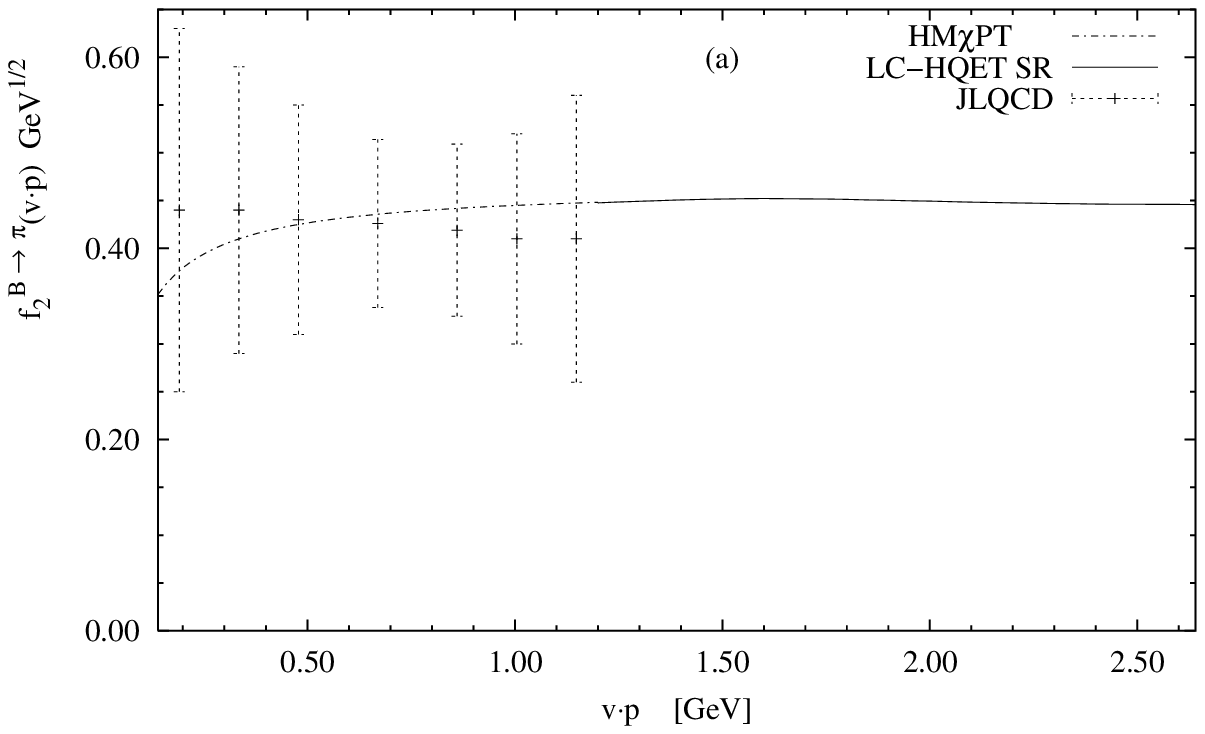} 
}
    \vspace*{3mm}
\centerline{
   \epsfysize=7.5cm \epsfxsize=12.5cm
   \epsfbox{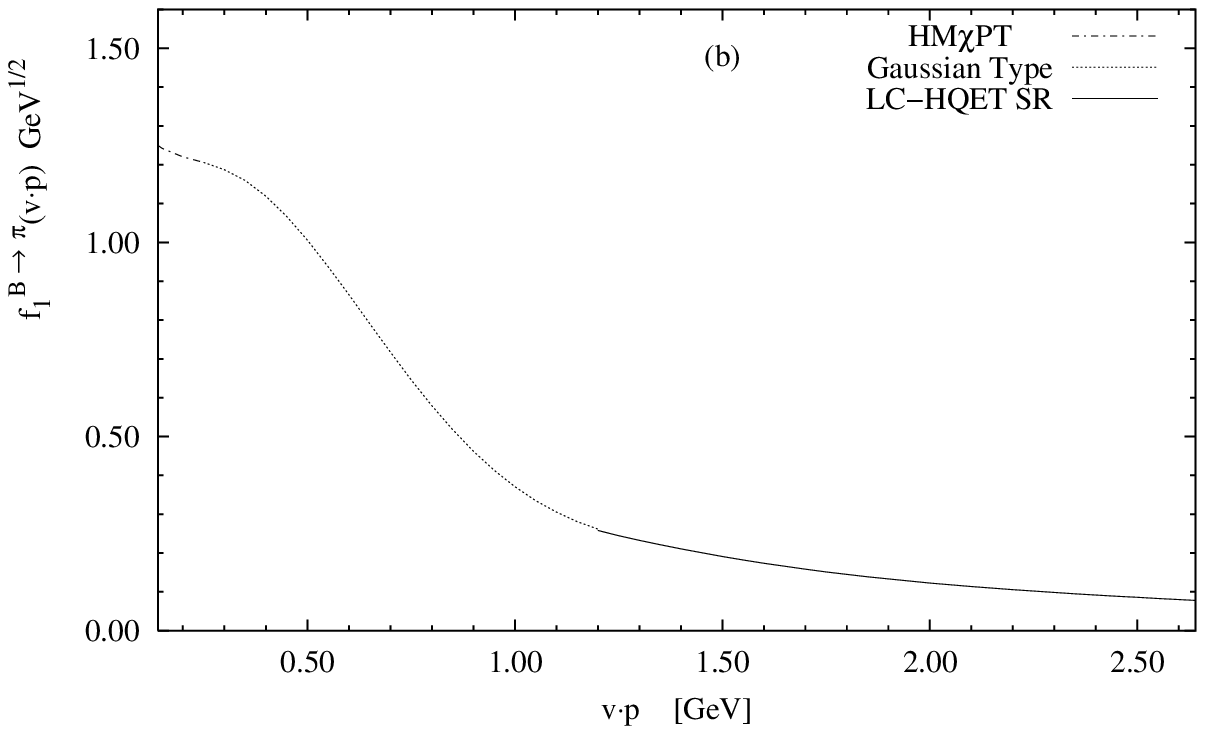}
}
\vspace*{-1mm}
\caption{The form factors $f^{B\to \pi}_{1,\,2}(v\cdot p)$ as obtained from 
LC-HQET sum rules including a matching to the HM$\chi$PT result in the soft
pion region.
\label{UF_all}}
\end{figure}

It is interesting to compare our results with those of other approaches.
First let us look at the soft pion limit.  The heavy meson
chiral perturbation theory (HM$\chi$PT)~\cite{Wise} (see also~\cite{Dominguez})
describes $f_1(v\cdot p)$ 
and $f_2(v\cdot p)$ in a single pole form in the soft pion region. Moreover, 
the $1/m_Q$ corrections do not change this behaviours~\cite{bpi,B_G}. In the 
case of $B\to \pi$ transitions, they can be expressed by
\beq
f_{1\,{}_{\rm HM\chi PT}}^{{B\to \pi}}(v\cdot p)&=& 
\frac{C_1 F}{2f_\pi}\,\Big(1+\frac{\delta F}{2m_b}\,\Big)\,
\Big[ \,1- \frac{g\;v\cdot p}{v\cdot p+\Delta_B}\,\Big]\,,  \\
f_{2\,{}_{\rm HM\chi PT}}^{{B\to \pi}}(v\cdot p)&=& 
\frac{C_1 F}{2f_\pi}\,\Big(1+\frac{\delta F}{2m_b}\,\Big)\,
\,\frac{g\; v\cdot p}{v\cdot p+\Delta_B}\,,
\eeq
where $\Delta_B\simeq m_{B^*}-m_B=0.045$ GeV, $g\sim 0.3$ is the coupling
of the pion to the heavy meson.
It is easy to see that our result for $f^{B\to \pi}_1(v\cdot p)$ does not 
match to the HM$\chi$PT result. Concerning $f^{B\to \pi}_2(v\cdot p)$ our
sum rule result Eq.~(\ref{sumruleB}) can be seen to reasonably well match
on to the soft pion result considering the uncertainties
in the values of $F$ and $g$. By taking $g=0.28$, one can find that the
extrapolation 
of HM$\chi$P to large $v\cdot p$ matches quite well with the sum rule 
calculations at intermediate pion energies as shown in Fig~\ref{UF_all}~(a).  
For practical purposes, we use the following Gaussian-type function to make a 
smooth connection between the sum rule result of $f^{B\to \pi}_1(v\cdot p)$ 
and that of HM$\chi$PT,
\beq  \label{Pole2}
f^{B\to \pi}_{1\,{\rm G}}(v\cdot p) &=& 
\frac{C_1 F}{2f_\pi}\,\Big(1+\frac{\delta F}{2m_b}\,\Big)\,
\Big[ \, g_0+g_1{\rm e}^{-g_2(v\cdot p-\delta_B)^2}\, \Big]\,,
\\[1mm]
&&~~~~(\,{\rm for ~~0.25 ~GeV}<v\cdot p < 1.2 ~{\rm GeV}\,)~ \nnb
\eeq
where $g_0=0.125$, $g_1=0.751$, $g_2=3.0$ GeV$^{-2}$ and $\delta_B=0.23$ GeV
were obtained by matching both sides. This is plotted in Fig.~\ref{UF_all}~(b).
It should be mentioned that the matching results obtained here correspond to 
the case of central values of $\omega_c$.  For further support of our matching 
procedure we have included in the lattice NRQCD results from the JLQCD 
collaboration~\cite{JLQCD}.  They can be seen to be well consistent with our 
matching result.

The large pion energy limit is another interesting limit to consider  
because the large energy effective theory (LEET) \cite{leet1,leet2} provides 
some model-independent information on this kinematical region.
In the $B\to\pi$ semileptonic decay, the light non-spectator 
quark gets a large amount of energy from the decay of the bottom quark.
The light spectator quark system interacts with the energetic quark mainly
at the energy scale $\Lambda_{\rm QCD}$ which is basically fixed by the size
of hadrons.  
New symmetries appear in the limit of $v\cdot p/\Lambda_{\rm QCD}\to\infty$. 
They are subject to corrections by hard gluon exchange \cite{leet3}. In the 
leading order of the heavy quark expansion our results are 
\begin{equation}
L_a(\infty) = 0\,, ~~~
L_b(\infty) = \frac{f_\pi}{F}T\phi_\pi(1)\,I_0\,,
\end{equation}
where 
\begin{equation}
I_i = \frac{1}{4}e^{2\bar{\Lambda}/T}\int_0^{\omega_c/T}{\rm d}x x^i e^{-x}\,,
\end{equation}
This agrees with LEET.  For the $1/m_Q$ corrections, our
 results are 
\begin{equation}
\begin{array}{lll}
F_6(\infty)       &=& \displaystyle -\frac{f_\pi}{F}\,
\frac{\mu_\pi}{6}\phi_\sigma(1)\,I_0\,,
\\[3mm]\nonumber 
\delta L_a(\infty)  &=& 0\,, \\[3mm]\nonumber
\delta L_b(\infty)  &=& \displaystyle \frac{f_\pi}{F}\,T \,
\Big[\,(2\delta\bar{\Lambda}/T-\delta F)\,I_0
-T\,I_1)\,\Big]\,\phi_\pi(1)\,.
\end{array}
\end{equation}
This can be compared with the LEET results \cite{leet4} only after the hard
gluon effects have been incorporated into these and can thus only been done
in the future. There even is a school which assumes that
the perturbative contribution is dominant for the $B\to\pi$ transition
\cite{leet5}. It is obvious that the LEET limit deserves further studies
which certainly will be done in the near future.

It is also interesting to compare our method and our results to the full QCD
LC sum rule calculation. The authors of Ref. \cite{KRW} in fact considered
the heavy quark limit of their full QCD LC sum rule calculation.
Their results differ from $L_a(v\cdot p)$ and $L_b(v\cdot p)$ only by a simple
transformation which were given in Eqs.~(47) and (46) in their paper. There
is a subtle difference from our results, though, which can be seen by
letting $\displaystyle{\chi (1-u)/u}\equiv\nu$.
It is only after taking $\nu/v\cdot p\rightarrow 0$ that the results of the
two calculations fully agree.
This limit may need further understanding. Numerically the QCD LC sum rules 
gave stable results for $q^2\leq 17$ GeV$^2$ which is consistent with the HQET 
LC sum rules requiring $v\cdot p \geq \omega_c/2\sim 1$ GeV.  These two 
methods could be in principle the same, provided that all the sub-leading 
corrections have been included in the calculations. Lacking  such powerful 
calculations, the effective theory calculates a physical quantity in a most 
thorough and least complicated way through clearly separating the perturbative 
and nonperturbative parts of the quantity.  Therefore in certain appropriate 
regions of the pion's phase space the HQET calculation may give more
reliable results. Practically speaking the method is to calculate form
factors at the hadronic scale, and the perturbative contribution is accounted
for by multiplying in renormalization factors.  

\begin{figure}[htb]
\vspace*{-0.9cm}
\centerline{
   \epsfysize=9.7cm \epsfxsize=14cm
   \epsfbox{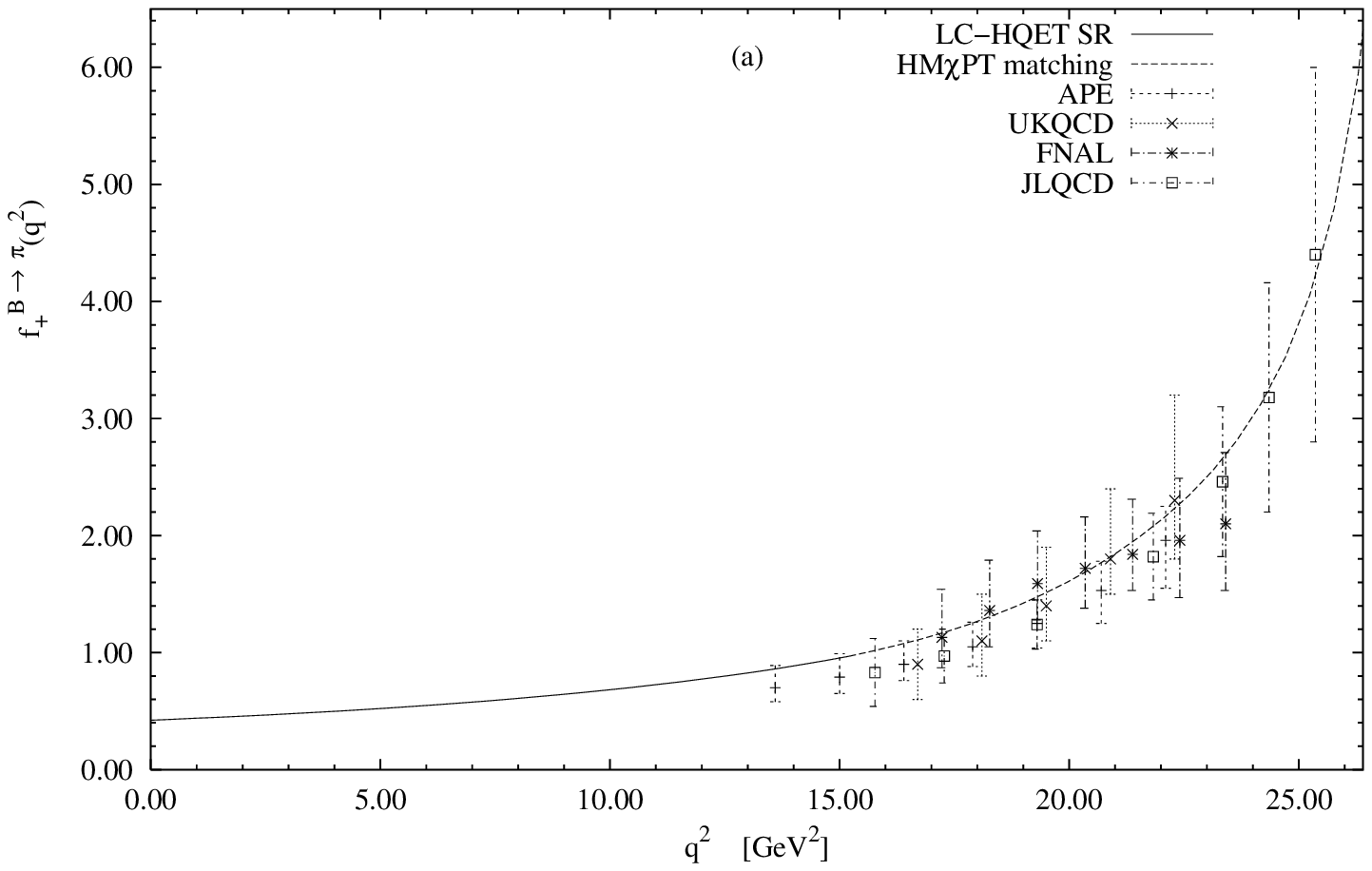} 
}
   \vspace*{-5mm}
\centerline{
   \epsfysize=9.7cm \epsfxsize=14cm
   \epsfbox{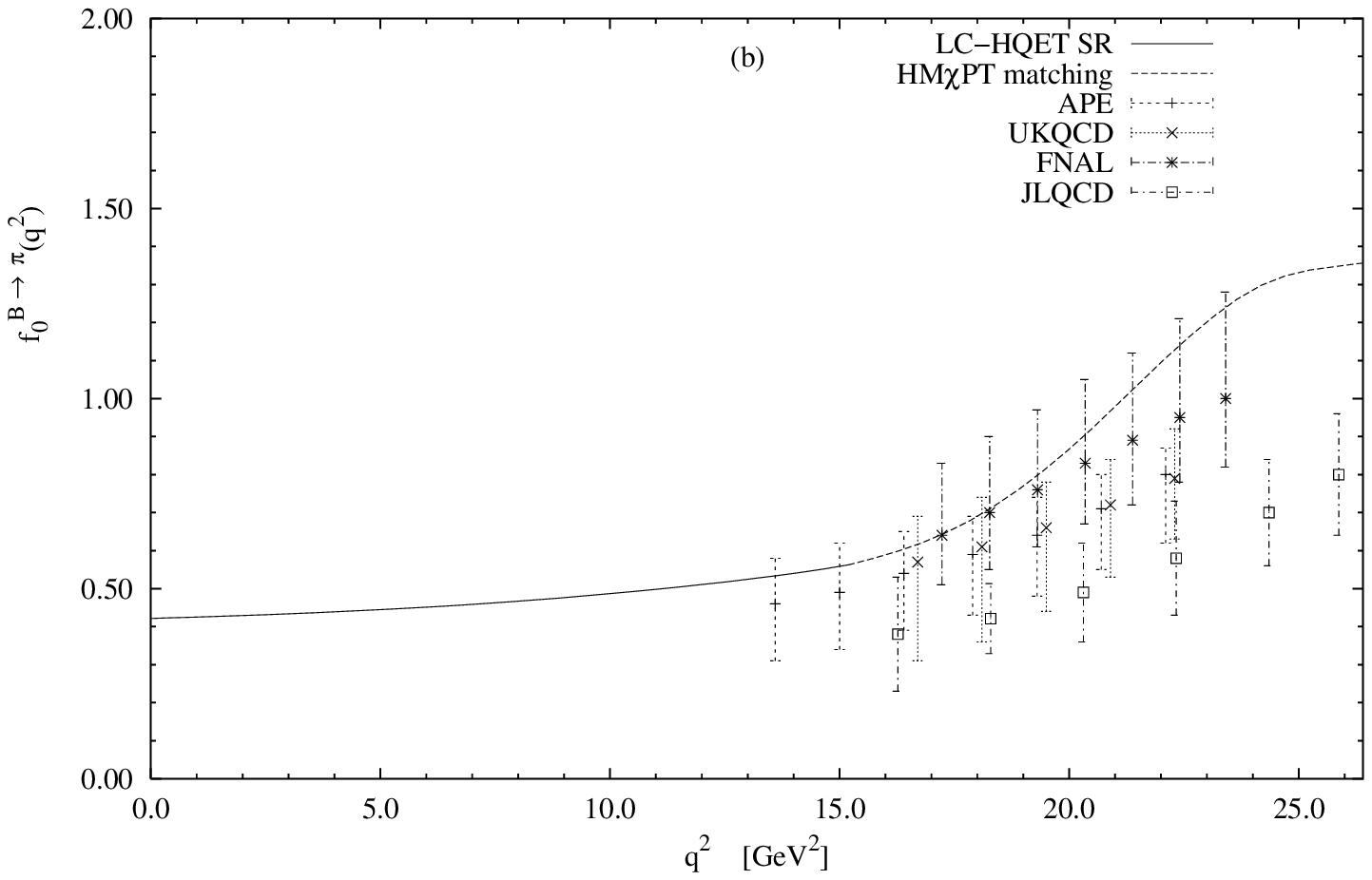}
}
\vspace*{-2mm}
\caption{The form factors $f^{B\to \pi}_{+,\,0}(q^2)$ from combining LC-HQET 
sum rules and HM$\chi$PT, and from the lattice QCD simulations by 
APE~{\cite{APE}}, UKQCD~\cite{UKQCD}, FNAL~\cite{FNAL} and JLQCD~\cite{JLQCD}.
\label{Fp0}}
\end{figure}

It is physically useful to reconstruct the conventional form factors defined
in Eq.~(\ref{FF}) from the results combining LC-HQET sum rules and HM$\chi$PT, 
by using the relation Eq.(\ref{FFR}). In Fig.~\ref{Fp0}~(a) we present 
$f^{B\to \pi}_{+}(q^2)$, which is directly measurable in semi-leptonic 
decay involving light leptons. In Fig.~\ref{Fp0}~(b) we present our result
for the scalar form factor 
$f^{B\to \pi}_{0}(q^2)$. The scalar form factor contributes to the decay
$B\to\pi\bar\tau\nu_{\tau}$
and also enters the factorized amplitudes in non-leptonic 
two-body $B$ decays.  For comparison, the 
results from the lattice QCD simulations by APE~{\cite{APE}}, 
UKQCD~\cite{UKQCD}, FNAL~\cite{FNAL} and JLQCD~\cite{JLQCD} are also shown in 
the figures.

In Table~\ref{tab:Tab} we compare our
results to the results of other model calculations which includes  
the conventional HQET QCD sum rule calculation of Ref.~\cite{Colangelo}.
We find
that our results for $f^{B\to \pi}_+(q^2)$ are 
about $14 \sim 33\%$ larger than those from the full QCD LC sum 
rule~\cite{lc,KRWWY}calculation which also includes $\alpha_s$ corrections.
\begin{table}[htb]
\caption{Comparison between different approaches for the universal functions
$L_a(v\cdot p)$ and $L_b(v\cdot p)$, and the form factor $f_+^{B\to \pi}(q^2)$
at some different kinematical points.
\vspace*{2mm} \label{tab:Tab}}
\tabcolsep 3mm
\begin{tabular}{@{} rccccc}
\hline
       & $L_a(v\cdot p)$ [GeV$^{1/2}$] & $L_b(v\cdot p)$ [GeV$^{1/2}$]
                                       & ~~ & $f_+(q^2)$ & ~~   \\
$v\cdot p$ [GeV] or $q^2$ [GeV$^2$] & $1.50$~~~~~~$2.64$ & $1.50$~~~~~~$2.64$
         & 0.0 & 6.0 & 12.0 \\
\hline
LC-HQET SR (NLO)  & ----- & ----- & $0.42\pm 0.04$ & 0.54 & 0.77   \\
LC-HQET SR (LO)   & 0.13~~~~~~0.06 & 0.34~~~~~~0.32 & 0.36 & 0.47 & 0.68 \\
HQET SR in LO~\cite{Colangelo} & 0.15~~~~~~0.13 & 0.25~~~~~~0.21 & 0.24 & 0.32
                         & ---   \\
LC-QCD SR~\cite{KRWWY,lc}  & ----- & ----- & $0.28\pm0.05$ & 0.42 & 0.66  \\
\hline
\end{tabular}
\end{table}
%

Finally, in order to be concrete, we give our prediction for the decay 
width of the decay $B^0\to \pi^- e^+\nu_e$ using our form factors.
We obtain
\beq
\Gamma &=& \left | \frac{V_{ub}}{4.08\times 10^{-3}} \right |^2 \,
(1.46\pm 0.30) \times 10^{-16}~{\rm GeV}\,.
\eeq
By taking $|V_{ub}|=(4.08\pm 1.18)\times 10^{-3}$ from inclusive measurement
of $B\to X_u \ell \nu$~\cite{NewExp}, we get
$\Gamma = (1.46\pm 0.30^{\,+0.97}_{\,-0.72})\times 10^{-16}$ GeV, where
the uncertainties are from that of form factors and $|V_{ub}|$ respectively.  
On the other hand, from the experimental result given in Ref.~\cite{CLEO96},
$\tau_{B^0}=(1.55\pm 0.03)$ ps$^{-1}$ and
Br$(B^0\to \pi^- e^+\nu_e)= (1.8\pm 0.6)\times 10^{-4}$,
we extract $|V_{ub}|=(2.94^{\,+0.33}_{\,-0.24}\pm 0.50)\times 10^{-3}$ with
the uncertainties from the form factors and the experiments, respectively.

Let us also comment on Ref. \cite{ww} which gave leading order results for 
$L_\alpha$.  In addition to that the analytical expressions of the sum rules 
differ from ours by a factor of $2$, the numerical input used in 
Ref. \cite{ww} is quite different, such as the choice of the energy scale 
$\mu$ and the determination of the sum rule window. This naturally affects the 
final results on $L_\alpha (v\cdot p)$.  

To summarize, we have applied the LC sum rule method to calculate the 
$B\to\pi\ell\nu$ weak decay form factors to order $1/m_Q$ in the framework of 
HQET.  We have calculated the leading and the relevant sub-leading universal 
form factors. Our form factor results have been matched to the appropiate soft 
pion results. We have also discussed the large pion energy limit of our 
results.  The full QCD LC sum rules and lattice QCD results have been compared 
with. In the future we are planning to include perturbative QCD corrections, 
try to incorporate the large energy effective theory into the QCD sum rule 
technique and perform detailed phenomenological analysis.

\section*{Acknowledgments}

We would like to thank Chao-Shang Huang, Stefan Groote and M. Beneke for 
helpful discussions. C.L. acknowledges support from the Alexander von Humboldt 
Foundation.  This work was supported in part by the National Natural Science 
Foundation of China with the grant no. 10075068, and by the BEPC National Lab 
Opening Project.  We would like to thank ICTP for hospitality where part of 
this work was done.


\end{document}